\def\conference{2} %

\if \conference1
    \documentclass[letterpaper,twocolumn,10pt]{article}
    \usepackage{usenix-2020-09}%
\fi

\if \conference3
    \documentclass[letterpaper,twocolumn,10pt,table]{article}
\fi

\if \conference2
    \documentclass[conference,compsoc]{IEEEtran}
\fi

\if \conference2
    \ifCLASSOPTIONcompsoc
      \usepackage[caption=false,font=footnotesize,labelfont=sf,textfont=sf]{subfig}
    \else
      \usepackage[caption=false,font=footnotesize]{subfig}
    \fi
\fi

\usepackage{tikz}
\usepackage{amsmath}
\usepackage{booktabs}
\usepackage{comment}
\usepackage{array}

\usepackage{relsize}
\if \conference1
    \usepackage{graphicx}
    \usepackage{titlesec}
    \usepackage[normalem]{ulem}

    \titleformat{\subsection}[hang]
    {\normalfont\bfseries}
    {\thesubsection}{0.5em}{}
    
    \titlespacing*{\subsection}{0pc}{0.5em}{0.5em}
    \titlespacing*{\subsubsection}{0pc}{0.5em}{0.5em}

\fi

\if \conference2
\usepackage[hyphens,spaces,obeyspaces]{url}
\usepackage[colorlinks = true,
            linkcolor = blue,
            urlcolor  = blue,
            citecolor = blue,
            anchorcolor = blue]{hyperref}         %
\usepackage[compact]{titlesec}
\fi

\usepackage{filecontents}
\usepackage{multirow}
\usepackage{pdflscape}
\usepackage{rotating}

\usepackage{enumitem}
\setlist{nolistsep}

\usepackage{cleveref}
\usepackage{listings}
\usepackage{svg}
\usepackage[normalem]{ulem}

\usepackage{cleveref}
\crefname{section}{§}{§§}
\Crefname{section}{§}{§§}

\titlespacing{\section}{0pt}{2ex}{1ex}
\titlespacing{\subsection}{0pt}{1ex}{1ex}
\titlespacing{\subsubsection}{0pt}{1ex}{0ex}
\titlespacing{\paragraph}{0pt}{1.25ex plus 1ex minus .2ex}{1em}

\newcolumntype{B}{>{\hsize=1.714\hsize}X}
\newcolumntype{M}{>{\hsize=0.708\hsize}X}
\newcolumntype{S}{>{\hsize=0.577\hsize}X}
\newcolumntype{P}{>{\hsize=0.7\hsize}X}
\newcolumntype{Q}{>{\hsize=1.3\hsize}X}
\newcolumntype{Y}{>{\centering\arraybackslash}X} %
\newcolumntype{L}{>{\raggedright\arraybackslash}p} %
\newcolumntype{C}{>{\centering\arraybackslash}p} %
\newcolumntype{R}{>{\raggedleft\arraybackslash}p} %

\if \conference2
\makeatletter

\Hy@AtBeginDocument{
	\def\@pdfborder{0 0 1} %
	\def\@pdfborderstyle{/S/U/W 0.5} %
}
\makeatother
\fi

\if 1\conference
    \usepackage[
        backend = biber,
        sorting = none, %
        citestyle = numeric-comp, %
        maxbibnames = 3, %
        mincrossrefs = 1000, %
        url=false
    ]{biblatex} 
\fi

\if \conference2
    \usepackage[
        backend = biber,
        sorting = none, %
        maxbibnames = 3, %
        mincrossrefs = 1000, %
        citestyle = numeric,
        url=false,
        doi=false,
    ]{biblatex} 
\fi

\if \conference1
    \usepackage[
        backend = biber,
        sorting = none, %
        citestyle = numeric-comp, %
        maxbibnames = 3, %
        mincrossrefs = 1000, %
    ]{biblatex} 
\fi

\DeclareSourcemap{
	\maps[datatype=bibtex, overwrite]{
		\map{
			\step[fieldset=editor, null]
			\step[fieldset=location, null]
			\step[fieldset=publisher, null]
			\step[fieldset=isbn, null]
			\step[fieldset=issn, null]
			\step[fieldset=pages, null]
		}
	}
}

\makeatletter
\newcommand{\starfootnote}[1]{%
  \begingroup
    \renewcommand{\thefootnote}{\fnsymbol{footnote}}%
    \let\IEEE@orig@makefntext\@makefntext
    \renewcommand{\@makefntext}[1]{%
      \parindent 1em%
      \noindent
      \hb@xt@1.8em{\hss\@thefnmark\ }##1%
    }%
    \footnotetext[1]{#1}%
    \let\@makefntext\IEEE@orig@makefntext
  \endgroup
}
\makeatother

\definecolor{codegreen}{rgb}{0,0.6,0}
\definecolor{codegray}{rgb}{0.5,0.5,0.5}
\definecolor{codepurple}{rgb}{0.58,0,0.82}
\definecolor{backcolour}{rgb}{0.95,0.95,0.92}

\lstdefinestyle{mystyle}{
  backgroundcolor=\color{backcolour}, commentstyle=\color{codegreen},
  keywordstyle=\color{magenta},
  numberstyle=\tiny\color{codegray},
  stringstyle=\color{codepurple},
  basicstyle=\ttfamily\footnotesize,
  breakatwhitespace=false,         
  breaklines=true,                 
  captionpos=b,                    
  keepspaces=true,                 
  numbers=left,                    
  numbersep=5pt,                  
  showspaces=false,                
  showstringspaces=false,
  showtabs=false,                  
  tabsize=2
}

\begin{document}

\date{}

\title{Learned, Lagged, LLM-splained: LLM Responses to End User Security Questions} %

\if\conference1
\fi

\if\conference2

\author{
    \IEEEauthorblockN{Vijay Prakash}
    \IEEEauthorblockA{New York University\\
    \small vp2179@nyu.edu}
    \and

    \IEEEauthorblockN{Kevin Lee}
    \IEEEauthorblockA{JPMorgan Chase\\
    \small kevin.lee@jpmorgan.com}
    \and

    \IEEEauthorblockN{Arkaprabha Bhattacharya~}
    \IEEEauthorblockA{Cornell Tech\\
    \small ab2956@cornell.edu}
    \and

    \IEEEauthorblockN{~Danny Yuxing Huang*}
    \IEEEauthorblockA{New York University\\
    \small dhuang@nyu.edu}
    \and

    \IEEEauthorblockN{Jessica Staddon*}
    \IEEEauthorblockA{Northeastern University\\
    \small j.staddon@northeastern.edu}
}

\fi

\maketitle

\renewcommand{\paragraph}[1]{\vspace{0.1cm} \noindent \textbf{#1}}

\def\diffmode{2}

\if\diffmode1
\newcommand{\add}[1]{{\color{ForestGreen} {#1}}}
\newcommand{\remove}[1]{{\color{red} {\sout{#1}}}}
\else
\newcommand{\remove}[1]{{\color{red}{}}}
\newcommand{\add}[1]{{\color{black} {#1}}}
\fi

\newcommand{\kevin}[1]{{[\color{brown} KL: {#1}]}}
\newcommand{\arka}[1]{{\color{orange} {#1}{-- arka}}}

\newcommand{\vijay}[1]{{\color{purple} {#1}{-- Vijay}}}
\newcommand{\jessica}[1]{{\color{blue} {#1}{-- jessica}}}
\newcommand{\danny}[1]{{[\color{cyan} DH: {#1}]}}

\newcommand*\circledsmall[1]{\tikz[baseline=(char.base)]{
            \node[shape=circle,draw,inner sep=1pt, scale=0.7] (char) {#1};}}
            
\newcommand*\circledmed[1]{\tikz[baseline=(char.base)]{
            \node[shape=circle,draw,inner sep=1pt, scale=0.8] (char) {#1};}}

\newcommand*\circledfilled[1]{\tikz[baseline=(char.base)]{
            \node[shape=circle, draw, fill=black, text=white, inner sep=1pt,] (char) {#1};}}

\newcommand*\circledmedfilled[1]{\tikz[baseline=(char.base)]{
            \node[shape=circle, draw, fill=black, text=white, inner sep=0pt, scale=0.8] (char) {#1};}}

\newcommand*\circledsmallfilled[1]{\tikz[baseline=(char.base)]{
            \node[shape=circle, draw, fill=black, text=white, inner sep=0pt, scale=0.7] (char) {#1};}}

\newcommand*\squaremed[1]{\tikz[baseline=(char.base)]{
            \node[shape=rectangle, draw, fill=black, text=white, inner sep=2pt, scale=0.8] (char) {\textbf{#1}};}}

\newcommand{\questionsCollected}{858 }
\newcommand{\questionsEvaluated}{900 }
\newcommand{\postIRRQuestions}{828 }
\newcommand{\geminiQuestionsEvaluated}{172 }
\newcommand{\llamaQuestionsEvaluated}{172 }
\newcommand{\failurepattenrcountLLaMAandGemini}{14}
\newcommand{\failurepattenrcountGPT}{13}

\newcommand{\actionableSuggestion}{Actionable suggestion}
\newcommand{\asShort}{
    \circledmedfilled{AS}
}
\newcommand{\devsuggestion}{
    \circledmedfilled{D}
}
\newcommand{\usersuggestion}{
    \circledmedfilled{U}
}

\newcommand{\inlinecircle}[2][1.2ex]{
    \tikz[baseline=(char.base)]{
        \node[shape=circle, fill=black, text=white, inner sep=0pt, minimum size=#1] (char) {#2};
    }
}

\newcommand{\ucircle}[1][1.2ex]{
    \inlinecircle[#1]{U}
}

\newcommand{\LackAdherenceResearch}{Failure to leverage research}
\newcommand{\Overprescription}{Over-prescription of security technologies and products}
\newcommand{\TransparencyArguments}{Missing transparency arguments that build trust among users}
\newcommand{\Permissions}{Listing used permissions and their Explanation}
\newcommand{\Scams}{Explaining scams and their trends}
\newcommand{\MakingUpFactsAV}{Making up facts about the AV technology}
\newcommand{\MissingThreatAngles}{Missing threat angles}
\newcommand{\Hallucinations}{Security misinformation due to hallucinations}
\newcommand{\TemporalDependency}{Temporal dependency unawareness}
\newcommand{\UISteps}{Incorrect and incomplete user interface (UI) steps}
\newcommand{\IncorrectURLsEmails}{Incorrect URLs and email addresses}
\newcommand{\ResponsesGearedTowardsDevelopers}{Responses geared towards developers and system administrators}
\newcommand{\MisinterpretQuestions}{Misinterpreting the questions}
\newcommand{\RestrictiveGuardrails}{Overly restrictive safety guardrails}
\newcommand{\IndirectResponses}{LLM-splaining}

\newcommand{\failureReason}{Reasons behind this error pattern: }

\renewenvironment{quote}
  {\list{}{\leftmargin=-0.0em%
            \topsep=0pt
            }\item[]\itshape\footnotesize}
{\endlist}

\begin{abstract}
Answering end user security questions is challenging. While large language models (LLMs) like GPT, Llama, and Gemini are far from error-free, they have shown promise in answering a variety of questions outside of security. We qualitatively evaluated responses from three popular LLMs to 900 systematically collected security questions in the first such evaluation in the area of end user security. %

While LLMs demonstrate broad generalist ``knowledge'' of end user security information, there are  
patterns of errors and limitations across LLMs---including stale, inaccurate, and incomplete answers, as well as indirect or unresponsive communication styles---which negatively impact the user experience. Based on these patterns, we suggest directions for improved model development and
recommend user strategies for interacting with LLMs when seeking assistance with security.
\end{abstract}

\def\showreview{2}

\if\showreview1
\newpage

\hrulefill

In \Cref{review}, we present the peer review of the submission from another conference. Our paper addressing reviewer comments starts from \Cref{sec:intro}, where removed contents are in red, and our updated responses are in green.

\hrulefill
\section{Peer Review Comments} \label{review}

\subsubsection*{Review \#357A}
\hrulefill

\paragraph{Paper Summary}
This paper examines the performance of LLMs in answering end-user security questions. It evaluated 3 LLMs on 900 security questions and found that while LLMs can answer general security questions, there are also errors and limitations. The paper made model improvement suggestions accordingly.

\paragraph{Technical Correctness}
\begin{itemize}
    \item Fixable Major Issues
\end{itemize}

\paragraph{Technical Correctness Comments}
I was not sure about the questions collection process. See below for details.

\paragraph{Scientific Contribution}
\begin{itemize}
    \item Addresses a Long-Known Issue
    \item Identifies an Impactful Vulnerability
\end{itemize}

\paragraph{Presentation}
\begin{itemize}
    \item No Flaws in Presentation
\end{itemize}

\paragraph{Comments to Authors}
There is a spike of papers that evaluate LLMs in different tasks. This particular paper focuses on security questions, which can be an interesting application domain as such questions may have a real-world impact on users’ security, safety, and potentially privacy as well. I have some concerns about the paper:
\begin{itemize}
    \item I found myself having a hard time understanding the process of question collection. It is not clear to me why the stated approach was chosen, given its several limitations. Security content based on trusted resources was not necessarily what users would ask. For example, a general user would have very few opportunities to work with network security, which typically requires more advanced knowledge. Additionally, the template only focused on “what” and “why” questions without addressing “how” questions, which most users might struggle with.
    \item There is also some confusion regarding the data analysis process. The paper once suggested that all four attributes were evaluated using a 3-point Likert scale but later mentioned different evaluation scales. The “thoroughness” attribute is particularly unclear—how was “completely achieved” determined?
    \item The idea of “reliability” is interesting but raises questions. For instance, how does paraphrasing influence the results? Were partial answers or repeated attempts considered in the study? It is unclear how meaningful the results related to reliability are, given the variability in responses.
\end{itemize}
Overall, I have fundamental concerns about the approach and would not argue for accepting this at this stage.

\paragraph{Recommended Decision}
Weak Reject (Can be Convinced by a Champion)

\paragraph{Reviewer Confidence}
Highly Confident

\paragraph{Should this submission be reviewed by the Research Ethics Committee?}
No

\hrulefill

\subsubsection*{Review \#357B}
\hrulefill

\paragraph{Paper Summary}
This research performs a qualitative study on LLM performance in the area of end-user security and online safety by evaluating three popular LLMs—GPT, LLaMA, and Gemini—on 900 systematically collected questions. The study categorizes errors into content and communication styles, identifying outdated guidance, omission of critical details, and misrepresentation of security technologies. The authors collected user questions on security and online safety covering seven security topics from frequently asked questions published by consumer safety and commercial organizations. Responses through zero-shot prompting are assessed for quality, specifically their accuracy, completeness, and relevance. Evaluation includes aspects of communication style, directness, and whether necessary and important information is presented first. The methodology of response evaluation uses spot-checks to ensure correctness and reliability checks through repeated question evaluation. The authors find that LLM responses are generally semantically similar, leading to a single response being generated for a given question and LLM.

\paragraph{Technical Correctness}
\begin{itemize}
    \item Fixable Major Issues
\end{itemize}

\paragraph{Technical Correctness Comments}
The research evaluates all 900 questions with GPT but limits the analysis of LLaMA and Gemini responses to only those where GPT failed. As a result, the cross-model comparison is incomplete and could be biased.

A detailed quantitative comparison between the three LLMs (e.g., accuracy scores across categories for LLaMA and Gemini) is missing.

The study implies that LLMs rely heavily on marketing materials and outdated information. Quantification of this impact would be beneficial, as would understanding the training data distribution.

It is unclear whether the user questions represent real-world user queries. The natural language and framing of actual end-user questions may differ significantly from the dataset.

The study uses 900 questions from multiple sources.

Ethical considerations and data privacy protocols are considered and addressed.

The plan to release the question corpus is important for reproducibility.

\paragraph{Scientific Contribution}
\begin{itemize}
    \item Provides a Valuable Step Forward in an Established Field
\end{itemize}

\paragraph{Scientific Contribution Comments}
This research evaluates LLMs' capabilities and limitations in addressing user security.

\paragraph{Presentation}
\begin{itemize}
    \item No Flaws in Presentation
\end{itemize}

\paragraph{Presentation Comments}
The paper is well-written, well-structured, and presented with examples and illustrations. The research addresses an important topic, with a corpus of systematically selected questions covering dominant security areas. Recommendations are provided to both users and developers.

\paragraph{Comments to Authors}
The quality of LLM responses to end user security questions has not been rigorously studied yet. Therefore, there is a critical need to evaluate the quality of LLM support in the user security content as adoption accelerates. In the beginning the paper discusses previous work and and says that there are already well-documented risks to relying on LLMs for information broadly, including hallucinations, challenges in reasoning, and vulnerability to concept drift.
Further, LLMs have been found to perpetuate known misconceptions in security and online safety.
It points out, that a larger and broader evaluation of LLM security responses (beyond known misconceptions) is needed in order to comprehensively outline the boundaries of LLM security “knowledge” to inform users about LLM shortcomings, and support future model development and improvement. The authors identify key insights and make recommendation:
\begin{itemize}
\item LLMs demonstrate broad generalist “knowledge” of end user security information 
\item there are patterns of errors and limitations across LLMs consisting of stale and inaccurate answers
\item indirect or unresponsive communication styles
\item paper suggests directions for model improvement 
\item and recommends user strategies for interacting with LLMs when seeking assistance with security
\end{itemize}

The authors focus only on security questions, excluding privacy-related queries, which may be even critical in the context of end users.

The study does not incorporate feedback from the users to understand the usability and effectiveness of LLM responses.

The study does not scale due to manual evaluation and also misses out to address scientific and systemic challenges by such an automated study. The limited size may also introduce bias into the results. Some explanations or intepretations are speculative, such as causes for deficiencies due to over-representation of marketing content in training datasets.

How do your recommendations generalize to other LLMs? and beyond user security?

\paragraph{Recommended Decision}
Weak Reject (Can be Convinced by a Champion)

\paragraph{Reviewer Confidence}
Fairly Confident

\paragraph{Should this submission be reviewed by the Research Ethics Committee?}
No

\newpage
\fi

\section{Introduction}
\label{sec:intro}
\starfootnote{ Both authors contributed equally as advisors.}

Large language models (LLMs) have grown increasingly popular among users for distilling complex information in user-friendly ways. This capability has made LLMs an appealing resource for computer security and online safety. For example, LLMs provide financial advice as chatbots \cite{next_web}, and are endorsed for fraud detection by security vendors \cite{zdnet_norton, biocatch, dataleon}. The use of LLMs for computer security advice is the natural evolution of the user practice of seeking online advice for security-relevant tasks such as strong password creation \cite{redmiles-2016-sources-n-selection}. However, the quality of LLM chat responses to end user security questions has not been rigorously studied.\footnote{Since we studied the role of LLMs solely as conversational assistants, we will refer to them as ``LLM'' throughout the remainder of the paper.} There is a critical need to evaluate the quality of LLM support in the user security context as adoption accelerates.

There are well-documented risks to relying on LLMs for information, including hallucinations, reasoning errors,
and vulnerability to concept drift~\cite{openai2024gpt4, GPT4SystemCard, geminiteam2024geminiv1, brown2020language, rajani-etal-2019-explain, shwartz-etal-2020-unsupervised, Lu_2018, bommasani2022opportunitiesrisksfoundationmodels}. Further, LLMs have been found to perpetuate known misconceptions in security and online safety~\cite{chen2023can}. A larger and broader evaluation of LLM security responses (beyond known misconceptions \cite{chen2023can}) is needed in order to help identify the boundaries of LLM security ``knowledge” and inform users and model developers of LLM shortcomings.%

To address this gap, we conducted the first expert-based manual evaluation of LLM responses to end-user security questions. We collected a diverse set of security and online safety questions from FAQs published by consumer safety and commercial organizations, covering seven key security areas. These questions are representative of common user concerns (\Cref{sec:represent}). We evaluated responses to these questions from three popular LLMs---GPT, Llama, and Gemini---using authoritative sources to assess their accuracy, completeness, and relevance. Our qualitative assessment focused on identifying patterns of errors and communication styles in LLM responses, providing insights into their strengths and limitations.

Our assessment is guided by the following research questions (RQs):

\begin{enumerate}[wide, label={\bfseries RQ\arabic*:}, left=0pt..0em, itemindent=2.6em]
    \item What is the information quality of LLM responses to user security questions? %
    \item What are the patterns of deficient or erroneous LLM responses to user security questions? %
    \item What actionable suggestions can improve LLM performance for users seeking security advice and developers building models?%
\end{enumerate}

We found that LLMs provide quality information expressed in a user-friendly manner when prompted with \textit{general knowledge} security questions. However, we also found consistent %
deficiencies and errors across LLMs. %
Surfacing these limitations can help users more effectively engage with LLMs and support improved model development. 

In particular, LLMs often fail to leverage research findings that could improve communication of important nuances in guidance and reduce misinterpretation of more specialized concepts. For example, LLMs provide outdated password guidance, neglect the risks of app permissions, fail to connect phishing and HTTPS URLs, and confuse zero-knowledge proofs and zero-knowledge encryption---all topics well-studied in research. 
In addition, our findings tentatively suggest LLMs may be influenced by marketing materials in that they tend to over-promise the capabilities of technology and products (e.g., endorsing VPNs for phishing protection), and miss important criticisms (e.g., neglecting the transparency benefits of open source code and data breach handling). %

We also find safety guardrails can interfere with answering benign security questions as can patterns of indirect communication style. %
Indirect LLM responses may begin  with encouraging or sycophantic \cite{sharma2023towards} statements (e.g., ``Great question!’’), or other information that has not been asked for and may already be known by the user (e.g., definitions). We term the latter ``LLM-splaining’’ given its similarity to mansplaining \cite{solnit2014men}.

\paragraph{Contributions.} We advance the understanding of LLM performance on user security questions through:
\begin{enumerate}[label={\bfseries \arabic*)}] %
    \item \textbf{A diverse set of LLM ``knowledge’’ %
   deficiencies in computer security}: Through inductive thematic analysis on 1244 responses across the 3 LLMs, we found multiple patterns of inaccuracy and omissions, some of which can lead to complete user-LLM interaction failures in which LLM responses are not relevant to the user's question. These patterns inform guidance for users and model developers (\Cref{findings},~\Cref{sec:discussion}). %
    \item \textbf{A corpus of user security questions with authoritative answer sources}: We curated and will publish a collection of user security questions from 7 categories with sources of authoritative answers for use by future research and development projects (\Cref{sec:method}).
    \item \textbf{A framework for assessing LLMs}: We adapted information integrity concepts to form a framework for the qualitative assessment of the content and communication style of LLM responses (\Cref{sec:method}).

    \item \add{\textbf{A corpus of expert-rated LLM responses:} $5$ security researchers conducted the first known expert manual evaluation of $1244$ LLM responses to user security and safety questions. The labeled corpus is potentially useful for training an automated LLM evaluation system. We will make the dataset available to researchers and developers upon request.}
\end{enumerate}

\section{Background and Related Work}
\label{sec:background}

\paragraph{Security and privacy advice for non-expert end users.}
Prior work has explored many aspects of user security and privacy advice, including the sources and behavioral impact of advice by demographic segment \cite{redmiles-2016-sources-n-selection, redmiles-2016-learned-to-be-secure}, user advice source selection \cite{redmiles-2016-learned-to-be-secure, pfeffer-2022-replication-infomal-lessons, emilee-2012-informal-lessons} and availability of advice \cite{sruti-2022-ambiguous-sec-advice}. The security mental models and misconceptions that advice would ideally address \cite{nicholson-2018-cybersurvival, herbert2023world} and how they are impacted by mass media \cite{fulton-2019-media-effect}, are also well-studied.

From user studies, the research community has found that expert security advice can be complex and difficult for users to follow \cite{reeder-2017-152-steps, redmiles2020webadvice}.
By conducting user studies involving users and administrators, Murray et al. found available security advice to be contradictory and ambiguous, resulting in user disagreement regarding the value of advice~\cite{murray-2023-costbenefitsofadvice}.

Closest to this paper is Chen et al.'s \cite{chen2023can} evaluation of LLMs' ability to refute security misconceptions, where they found Bard and ChatGPT correctly negate security misconceptions approximately $70\%$ of the time. To the best of our knowledge, this paper is only the second to consider the quality of LLM user security advice and the first to broadly assess LLM advice across security topics. Unlike Chen et al., who focused on a fixed set of known misconceptions and binary “Yes/No" responses, our evaluation considers open-ended user queries and assesses both the content and communication style of responses. While the scope of our work is different and broader, we do find evidence of some of the misconceptions they study, i.e., ``Websites that use HTTPS are trustworthy,'' and “HTTPS protocol could protect against phishing”. %

\begin{table}[!h]
    \relsize{-1}
    \centering
    \vspace{-1.0em}
    \setlength{\abovecaptionskip}{0pt}
    \caption{This paper and existing security advice research.}
    
    \resizebox{\linewidth}{!}{%
    \begin{tabular}{ p{0.60\linewidth} p{0.4\linewidth}}
        \toprule
        \textbf{Data source} & \textbf{Prior works}\\ 
        \midrule
        Interviews and questionnaires & \cite{redmiles2020webadvice, redmiles-2016-learned-to-be-secure, redmiles-2016-sources-n-selection, murray-2023-costbenefitsofadvice, robert-2017-152steps, emilee-2012-informal-lessons, pfeffer-2022-replication-infomal-lessons, ion-2015-expert-v-non-expert, busse-2019-replication-expert-v-non-expert} \\
        Q\&A websites & \cite{hasegawa-2022-non-experts-snp-qna} \\
        Mass media, and social media & \cite{fulton-2019-media-effect, sruti-2022-ambiguous-sec-advice} \\
        LLM responses & \cite{chen2023can}, this paper\\
        \bottomrule
    \end{tabular}
    }
    \label{sec-advice-related-work}
    \vspace{-1.0em}
\end{table}

\paragraph{LLM proliferation.} The adoption of LLMs has grown so swiftly that there is speculation they will replace traditional search engines \cite{pcmag-chatGPT-v-Google}. The LLMs we have evaluated, GPT-4 \cite{openai2024gpt4}, Gemini \cite{geminiteam2024geminiv1.5}, and Llama-3 \cite{Llama3git}, are used by a large number of users \cite{Liu-ai-usage-2024} as browser-based conversation bots offered by model developers: OpenAI uses GPT-based models in ChatGPT \cite{GPT4Research}; Meta uses Llama in its chatbot \cite{MetaLlama3launchblog,}; and Google uses Gemini in its chatbot of the same name \cite{GeminiReleaseBlog} and on the search page. Many other companies use models' APIs for customer support bots interfacing with users \cite{openaistories}.

\paragraph{LLM evaluations.} Safety and security evaluation by model developers has not covered the evaluation of user security questions \cite{openai2024gpt4, GPT4SystemCard}, \cite{geminiteam2024geminiv1.5, geminiteam2024geminiv1}. Indeed much of the security evaluation focus has been on the generation of harmful content (e.g., \cite{inan2023Llama}), and insecure code generation by LLMs and their ability to assist in cyberattacks in an automated way \cite{PurpleLlama, bhatt2023purple}.%

The machine learning research community has developed several LLM benchmarks \cite{chang-llms-eval-survey-2024}. To the best of our knowledge, the only benchmarks overlapping with computer security and privacy are: \cite{zhang-etal-2024-safetybench} that focuses on various aspects of online privacy practices (in addition to other topics) %
and, \cite{hendrycks2021measuring} that covers cryptography and other technical security topics.  
Questions in \cite{hendrycks2021measuring} and \cite{zhang-etal-2024-safetybench} are multiple-choice questions, whereas our questions are open-ended; \cite{mt-bench-zheng-2023} covers open-ended questions but doesn't include end-user security questions.

In contrast to these benchmarks, we provide the first prompt corpus and associated LLM responses to questions representing end-user security concerns across 7 security topics.

\if \conference1
\section{Methodology} \label{sec:method}
\fi
\if \conference2
\section{Method}
\label{sec:method}
\fi
We conducted a qualitative analysis of 900 security questions, focusing on errors and knowledge deficiencies across three LLMs. This evaluation establishes a baseline for LLM knowledge in security and safety, excluding privacy questions and large-scale quantitative comparisons.

We answered the RQs through three tasks. First, we built a corpus of questions reflecting end-user security concerns. Then, we collected responses to the questions from three LLMs in a ``zero-shot''~\cite{kojima2023large}, single-turn setting. Finally, we assessed the information quality and directness of the LLM responses. In this section, we describe how we implemented these tasks to prioritize common end-user questions and reliably measure LLM performance on the questions.

\subsection{Question Collection} 
\label{non-exp-question-collection} 

\begin{table}[!htb]

    \centering
    \caption{The security categories we identified and underlying topics}

    \begin{tabular}{L{0.13\textwidth}L{0.27\textwidth}}
    \toprule
        \textbf{Category} & \textbf{Topics} \\ 
    \midrule
Authentication & Strong passwords, Password managers, 2FA, Account recovery  \\ 
Social Engineering & Phishing, Scam  \\ 
Software Updates & Patches/Security Updates, General Software Updates, Outdated software, Device/Firmware Updates\\ 
Antivirus / Antimalware & Guidance for Identified Compromises, Running AV Scans, AV Capabilities and Rationale, Specific AV knowledge \\ 
Safe Browsing & E-Commerce, Internet Safety Rules, Types of Compromise, Compromised / Unsafe Webpages, Browser Security Settings  \\ 
Secure Wi-Fi / Network & Public Wi-Fi, Home Network, Bluetooth, NFC, VPN \\ 
Smart Devices Security & Smart home, Smartphones  \\ 
    \bottomrule
    \end{tabular}
    \label{tab:categories-table}
    \vspace{-1em}
    
\end{table}

\paragraph{Identifying security topics.} To answer our RQs, we built a corpus that would closely represent common end user security concerns. To start, we collected security topics from trusted security content resources made available to users from the National Institute of Standards and Technology (NIST) \cite{NIST}, National Cyber Security Center (NCSC) \cite{NCSC}, Cybersecurity and Infrastructure Security Agency (CISA)~\cite{CISA}, and National Cybersecurity Alliance (NCA) \cite{NCA}. We identified seven high-level security categories: Authentication, Social Engineering, Software Updates, Antivirus / Antimalware, Safe Browsing, Secure Wi-Fi / Network, and Smart Devices Security. We also noted the underlying fine-grained security topics that were discussed in these resources, shown in~\Cref{tab:categories-table}.

\add{\paragraph{Collecting questions.} We combined three different approaches to collect questions in the seven security categories. First, we developed Google Search queries related to topics, similar to the method Redmiles et al. used~\cite{redmiles2020webadvice}. With the queries, we identified websites meeting our trusted resource criteria that were popular sources of security guidance and mined them for FAQ guides. We defined trusted resources as organizations that either have a mission tied to user security (e.g., government agencies, universities) or a business that strongly depends on customer security (e.g., financial services, security vendors).\footnote{Our trusted resources collection includes security vendors, government agencies, non-profit organizations, financial institutions, university cybersecurity guides, and product websites (e.g., social media, browsers, and email providers). Some examples of trusted resources are the NIST, NCSC, CISA, NCA, Canadian Centre for Cyber Security (CCCS), Federal Communications Commission (FCC), Federal Trade Commission (FTC) and the Electronic Frontier Foundation (EFF).} %
Second, we included product sites related to the categories (Chrome, Edge, Facebook, Gmail, etc.) to collect security FAQs that are relevant to users. 
Third, we used templates to construct some questions related to the seven categories, such as ``What is [insert security concept, technology or product]?’’,  “Why should I trust [password manager] with my passwords?", “How to install [popular password manager]?", and ``What is a [type of scam]?" 
We collected questions from more than $50$ trusted resources in our corpus. We later evaluated the representation of actual user queries in these questions in~\Cref{sec:represent}, and also as part of our check on LLM response reliability in~\Cref{appendix-prompt-sensitivity}.

We found corpora from related research to be limited in scope~\cite{chen2023can,pattnaik2023perspectives,pattnaik2023perspectives} or were unable to be shared~\cite{redmiles2020webadvice}. Therefore, we built our own corpus of questions by mining dominant security topics and using them alongside question templates to extract user questions from {trusted resources}.

\paragraph{Authoritative sources.}
To assess the accuracy and completeness of LLM responses, we identified authoritative sources for each question in our study. These sources were selected based on established criteria for trusted resources and typically included FAQs, advice bulletins, or official documentation published by reputable organizations. For straightforward questions (e.g., “What is malware?”), a single authoritative source---the source of the question---often sufficed. However, more complex questions (e.g., “What are some other ways of defending against a phishing attack?”) required consultation of multiple sources and expert interpretation. On average, 1.7 authoritative sources were used per question. All sources were reviewed and validated for reliability by cybersecurity subject matter experts at a major financial services institution.
}

To classify the questions by the type of security information sought, two authors agreed that each question is most aligned with one of the following areas: \textbf{glossary and facts, guidance and best practices, attacks, trends, and risk assessment}. We also classified questions by knowledge category, \textbf{factual}, \textbf{conceptual}, and \textbf{procedural}, as in Bloom's taxonomy \cite{blooms-taxonomy-2001}. \add{See \Cref{tab:appendix-gpt-evaluation-dist-big-with-directness} for the number of questions in each category.} We will make the questions and their categorizations available for other researchers.

\subsection{Response Generation} \label{response-generation}
Prompt engineering is an active research area and recent studies suggest it is at best an emerging practice for users (e.g., \cite{zamfirescu2023johnny}), so we chose the default LLM conversation templates as the best available proxy for a typical user-LLM interaction. With the default templates, we
appended our questions to generate responses from GPT, Gemini, and Llama. 
The exact conversation template used to generate responses for all of our evaluations is below:
\begin{lstlisting}[language=Python]
conversation_messages = [
 {"role": "system",
  "content": "You are a helpful assistant."},
 {"role": "user",
  "content": "Hello!"},
 {"role": "assistant",
  "content":"Hello! How can I assist you today?"},
 {"role": "user", 
  "content": "-> Questions go here. <-"}]
\end{lstlisting}

\subsection{Response Evaluation} \label{evaluation-criteria}

\subsubsection{Evaluation Criteria \& Codebook}
We carefully reviewed LLM responses, assessing both content and communication style. With respect to content, we focus on information integrity and evaluate the accuracy, completeness, and relevancy of LLM responses as in Flowerday et al. \cite{flowerday-2007}. We also consider communication style by way of directness (also known as the inverted pyramid style of writing \cite{zinsser2001writing}), where necessary and important information is presented first. This communication style is often recommended for web \cite{Schade2018, Nielsen1996, Nielsen1997} and nonfiction writing. %
For our study, we define evaluation criteria as the following:

\begin{enumerate}
    \setlength\itemsep{0em}
    \item \textbf{Accuracy}: The proportion of verifiably correct information %
    \item \textbf{Thoroughness}: The degree to which all necessary information is provided%
    \item \textbf{Relevancy}: The proportion of information that is relevant%
    \item \textbf{Directness}: Whether the initial text is responsive to the question asked. %
\end{enumerate}

\remove{
Since there was no established codebook for interpreting these attributes in the user security context, we developed a
codebook (to be made available) that enabled the authors to consistently assess these attributes on a 3-point Likert scale. Each author coded LLM responses manually to evaluate the 4 attributes. %
The authors used information present in the authoritative sources of the question, along with their subject matter expertise, to assess all the facts in a response by following the detailed definition of four earlier mentioned criteria to the best of their ability.

In our codebook, the 3-point Likert scale for accuracy, thoroughness, and relevancy is summarized as: the attribute is completely achieved (e.g., in the case of accuracy, all information is determined to be correct), the attribute is partially achieved, and the attribute
is only minimally achieved. Directness is assessed as a binary label, i.e., the initial LLM text is directly responsive to the question asked or it is not. The authors also added notes explaining their evaluation of the attributes; these notes facilitated the analysis in \Cref{overall-evaluation}.}

\add{
Since there was no established codebook for interpreting these attributes in the user security context, we developed a codebook (to be made available) that enabled the authors to consistently assess these attributes. Each author coded LLM responses manually to evaluate the four attributes. The authors used information present in the authoritative sources of the question, along with their subject matter expertise, to assess all the facts in a response by following the detailed definition of four earlier mentioned criteria to the best of their ability.

In our codebook, for accuracy, thoroughness, and relevancy, a 3-point Likert scale is used: the attribute is completely achieved (e.g., in the case of accuracy, all information is determined to be correct), the attribute is partially achieved, and the attribute is only minimally achieved. Directness is assessed as a binary label, i.e., the initial LLM text is directly responsive to the question asked or not.

For thoroughness, the authors collectively decided and agreed upon the authoritative sources for each question. If an LLM response covers all aspects mentioned across the sources, then we rate it as ``completely thorough''. For a question about a specific platform or product, only one authoritative source is needed, which is the official documentation. For example, for the question “How to setup the Google Authenticator app?", we only used official Google documentation \cite{GoogleAuthenticatorSetup} %
on the other hand, for a question like “Are public Wi-Fi Networks Safe?", we used multiple authoritative sources from FTC \cite{FTC2024-public-wifi} %
and NSA \cite{NSA-wireless}.%
Furthermore, even though we use a 3-point Likert scale, we manually write down notes on what a particular LLM response is not perfect (e.g., not completely accurate, or not completely thorough). These notes facilitated the analysis in~\Cref{overall-evaluation}.}%

Examples of coded responses and the complete definition of each code point and its options in our codebook can be accessed \href{https://docs.google.com/document/d/e/2PACX-1vQ99tup4hftyJMDTYlRgL41AM560v-fFqWfSQzb2-G1emV-j_etghUB0-DYL0j9cMSQX7AYjF9d5Ojt/pub}{\textbf{here}}~\cite{codebook}.

Our attributes are motivated by the need to manage risk in AI systems and past research. For example, a
correct but incomplete answer could create user risk. %
Relevancy is important given user difficulty with prioritizing advice \cite{redmiles2020webadvice} and reconciling contradictions \cite{murray-2023-costbenefitsofadvice}. %
Finally, directness reduces the risk of losing user attention due to unnecessary information \cite{Nielsen1997}. %

\subsubsection{Coding the responses} %
\label{irr-establishment}

We qualitatively coded the responses on our evaluation criteria to measure quality, dividing the workload among four members of the research team. 
First, each coder coded 72 GPT responses each for consistency. Inter-rater reliability was calculated (Fleiss’s $\kappa$: Accuracy 0.73, Thoroughness 0.63, Directness 0.76) after three rounds~\cite{mchugh2012-irr}. There was 100\% agreement with Relevancy. Due to the substantial level of agreement, we divided the rest of the question responses among the coders for evaluation.

\paragraph{Breakdown of the 72 responses used to establish IRR.} %
Out of \questionsCollected collected questions, 30 were used during IRR. 10 questions out of these 30 were rephrased with Google search APIs (more details are in the prompt sensitivity experiment \Cref{appendix-prompt-sensitivity}) to generate another set of 42 semantically similar questions, totaling $72\ (=10 + 42 + 20)$.

\paragraph{Post-IRR evaluation of \postIRRQuestions responses.} In our dataset, LLM responses to open-ended questions contain $\sim$335 words on average. Evaluation of \postIRRQuestions responses of that size is a large amount of work, so we onboarded a fifth researcher to help code responses. The fifth researcher was trained to code responses using our codebook on some of the 72 responses the four researchers had coded during the IRR establishment phase, and the remaining were used to measure the IRR among 5 coders. After onboarding the 5th coder, the remaining \postIRRQuestions questions were split among the five researchers for evaluation.

 \paragraph{Total number of responses assessed and final evaluation assigned to them.} In our study, we evaluated \questionsEvaluated GPT responses in total, 72 during the IRR and 828 post-IRR. During the IRR establishment, coders did not have 100\% agreement in their evaluation of some responses, so to assign a final rating, these coders regrouped to discuss and assign a final collaborative rating. After dividing the rest of the questions, if a coder felt unsure about their assessment, they collaborated with another coder. In addition, every coder's evaluation was spot-checked by a second researcher to ensure that the original assessment was correct.

\subsection{LLM Response Reliability}
\label{sec:reproducibility}

Before making conclusions about LLM performance, it is necessary to gauge response reliability over multiple invocations of the same, or semantically similar, prompts. In \cite{chen2023can} some unreliability was found in responses to repeated and paraphrased queries. To the best of our knowledge, LLM response reliability has not been studied for open-ended user security questions, so we conducted two analyses to gauge response reliability for end-user security questions.

\subsubsection{Reliability for Paraphrased Questions} \label{appendix-prompt-sensitivity}

Our methodology to measure reliability in LLM responses across paraphrased questions (referred to as ``queries'' in \cite{chen2023can}) was the following:

\begin{enumerate}[wide, label={\bfseries \arabic*)}, left=0pt..0pt, itemindent=10pt]

    \item \textbf{Question selection:} First, we sampled 16 questions from our list of \questionsCollected questions such that they covered all the security categories in \Cref{tab:categories-table} and security question themes mentioned in the \Cref{non-exp-question-collection}.
    
    \item \textbf{Rephrasing questions using user queries:} We paraphrased original questions (OQs) using Google's autocomplete API \cite{google-auto-complete-api} (used by Google in the search bar) and related question suggestions \cite{google-query-rec} (often presented on the Google search result page as “People also ask", “More to ask", or “Questions related to your search"). Using these Google APIs, 16 questions were paraphrased 274 times, referred to as paraphrased questions (PQs) here. Then, using the transformer-based model Universal Semantic Encoder (USE) \cite{tensorflow-USE}, we measured how semantically close a Google API paraphrase is to the OQ to reduce the list of PQs to 157 for 15 OQs. Finally, 4 researchers manually verified whether the remaining 157 PQs for the 15 OQs were good paraphrases. At the conclusion, we had 10 OQs paraphrased to 42 questions (PQs). The number of times each OQ was rephrased ranged from 1 to 13. \add{To better represent organic user queries, we rephrase using Google Autocomplete that reflects user queries on the Google search engine \cite{google-autocomplete-2024}}. %

    \item \textbf{Response generation:} We collected responses for all 10 OQs and their 42 PQs by presenting them to an LLM model as prompts with the setup described in \ref{response-generation}.
    
    \item \textbf{Manual response similarity assessment:} Each pair of responses (one for an OQ and another for its PQ) was coded by two authors independently on a similarity level of the ordinal scale \textbf{completely similar}, \textbf{mostly similar}, \textbf{insufficiently similar}. Authors assigned similarity levels %
    following the guidelines described in our codebook \cite{codebook}. %
    Any disagreements between the two authors' assigned similarity levels were resolved through discussion, and a single similarity level was assigned.
    
\end{enumerate}

\subsubsection{Reliability for Repeated Questions} \label{appendix-repetive-consistency}

Our methodology for measuring consistency in LLM responses across repeated questions was the following: \textit{(\romannumeral 1)}\ We sampled 7 questions from our list of \questionsCollected questions; \textit{(\romannumeral 2)}\ Each question was presented to an LLM model 5 times to collect their response; \textit{(\romannumeral 3)}\ All 5 responses for a question were first individually rated by two authors on our evaluation criteria mentioned in \ref{evaluation-criteria} following the guidelines mentioned in our codebook \cite{codebook};
\textit{(\romannumeral 4)}\ Finally, for each question, authors rated 5 responses as a group on the similarity level of the ordinal scale \textbf{completely similar}, \textbf{mostly similar}, \textbf{insufficiently similar}.

\remove{\paragraph{Conclusion:} To gauge response reliability for user security questions, we gathered LLM responses to repeated and paraphrased questions and two authors manually reviewed them to assess the semantic similarity of the responses. We found LLM responses were generally semantically similar and consequently, we generated a single response for a given question and LLM in the remainder of the evaluation. See the detailed results of our evaluation in \Cref{appendix-repro-validation}.}

\add{\paragraph{Conclusion:} To gauge response reliability for user security questions, we gathered LLM responses to repeated and paraphrased questions and two authors manually reviewed them to assess the semantic similarity of the responses. We found LLM responses were mostly semantically similar and speculated that our results would be unlikely to change with actual user questions. Consequently, we generated a single response for each question and LLM, and used the questions collected from FAQs, opting not to include more questions in users' voices in the remainder of the evaluation. See the detailed results of our evaluation in \Cref{appendix-repro-validation}.}

\subsection{Error Pattern Discovery in LLM Responses} \label{overall-evaluation}

\paragraph{Qualitative Analysis of Imperfect GPT Responses.} In our qualitative analysis of imperfect GPT responses, we first conducted a manual review to identify recurring errors—specifically, mistakes or knowledge gaps present across multiple outputs. To catalog these errors systematically, we employed an inductive thematic analysis \cite{braun-2006-thematic-analysis}. We revisited every response that failed to satisfy at least one of our four evaluation criteria—accuracy, thoroughness, relevance, and directness—and assigned each a concise error label based on the coder’s notes. We then examined these labels and merged any that described similar error patterns, discarding labels that did not recur.

\paragraph{Evaluation of Llama and Gemini.} Next, to determine whether the error patterns we identified in GPT generalized to other models, we applied the same coding scheme to responses generated by Llama and Gemini. Given the considerable time required for manual evaluation, we limited our assessment to the subset of \geminiQuestionsEvaluated questions for which GPT had produced imperfect responses. For each of these items, we evaluated the corresponding Llama and Gemini outputs using our established codebook \cite{codebook} and then conducted a thematic analysis of their errors. After coding 304 responses (across both models), we observed that every error pattern previously attributed to GPT also appeared in Llama and Gemini, at which point we concluded our analysis.

Although our selective approach may have overlooked additional model-specific errors in Llama and Gemini, it nonetheless confirms that the error patterns we identified are shared across all three systems. Due to time constraints, a comprehensive exploration of any remaining issues in Llama and Gemini is left for future work.

The evaluation results of GPT for all the questions and the tags assigned during the error pattern discovery process are summarized in \Cref{findings}.%

\section{Findings} %
\label{findings}

\begin{table}[!h]
\if \conference1
\relsize{-1}
\fi
    \setlength{\abovecaptionskip}{0pt}
\caption{Performance of GPT in terms of \% of responses where it fully achieved the indicated attribute.}
\label{tab:gpt-evaluation-dist-by-topics}
\resizebox{\linewidth}{!}{%
    {\renewcommand{\arraystretch}{1.2}%
    \begin{tabular}{p{0.17\linewidth}p{0.37\linewidth}p{0.09\linewidth}p{0.16\linewidth}p{0.1\linewidth}p{0.11\linewidth}}
    \toprule
    \textbf{Categorization}& \textbf{Category} & \textbf{Accuracy} & \textbf{Thoroughness} & \textbf{Relevancy} & \textbf{Directness} \\
    \cline{3-6}
     & & Correct & Thorough & Relevant & Direct \\
    \midrule
    \textbf{} & \textbf{\# of responses out of all (900)} & 0.73 & 0.68 & 0.98 & 0.83 \\
    \cline{1-6}
    \multirow[t]{7}{\linewidth}{\textbf{Security Category}} & \textbf{Authentication (449)} & 0.69 & 0.69 & 0.98 & 0.83 \\
    \textbf{} & \textbf{Scams (83)} & 0.82 & 0.70 & 1.00 & 0.83 \\
    \textbf{} & \textbf{Safe browsing (50)} & 0.80 & 0.68 & 0.96 & 0.88 \\
    \textbf{} & \textbf{Antivirus (163)} & 0.61 & 0.59 & 0.99 & 0.86 \\
    \textbf{} & \textbf{Secure network/WiFi (46)} & 0.87 & 0.67 & 0.96 & 0.70 \\
    \textbf{} & \textbf{Smart devices (19)} & 0.89 & 0.63 & 0.95 & 0.79 \\
    \textbf{} & \textbf{Updates (90)} & 0.88 & 0.81 & 0.99 & 0.84 \\
    \cline{1-6}
    \multirow[t]{3}{\linewidth}{\textbf{Knowledge Category}} & \textbf{Factual (437)} & 0.76 & 0.72 & 0.98 & 0.84 \\
    \textbf{} & \textbf{Conceptual (164)} & 0.72 & 0.77 & 0.98 & 0.89 \\
    \textbf{} & \textbf{Procedural (299)} & 0.69 & 0.57 & 0.99 & 0.78 \\
    \bottomrule
    \end{tabular}
    }
}
\end{table}

In our evaluation, GPT, Llama, and Gemini attempted to answer almost all the questions in our corpus, however, often with errors in correctness, thoroughness, or relevancy. In this section, we briefly discuss the evaluation results, then \add{generalizable} error patterns seen across the three models in more detail, along with cause hypotheses. We discuss guidance for users and developers based on these patterns in \Cref{sec:discussion}.

As mentioned earlier, assessing the information integrity of LLM responses was time-consuming. Over 6 %
months, we evaluated 900 GPT responses (including establishing a high IRR). We present the findings of our GPT evaluation in \Cref{tab:gpt-evaluation-dist-by-topics}. Out of these 900 GPT responses, 415 were perfect (i.e., correct, thorough, relevant, and direct), and 485 were imperfect. GPT responses were accurate, thorough, relevant, and direct 73\%, 68\%, 98\%, and 83\% of the time, respectively. We report the detailed breakdown across different question categorizations in \Cref{tab:appendix-gpt-evaluation-dist-big-with-directness} in \Cref{gpt-eval-results}. \remove{We do not report the Llama and Gemini evaluation numbers because they were only evaluated for questions where GPT had failed, and reporting them could lead to incorrect LLM performance comparisons.}

\add{\paragraph{Note:} We do not report quantitative evaluation results for Llama and Gemini, as their responses were only assessed for a subset of questions where GPT had previously failed, see \Cref{overall-evaluation}. Reporting these selective results could lead to misleading comparisons of overall model performance.}

\paragraph{GPT evaluation overview:} First, there is a downward trend in GPT's accuracy from factual (highest accuracy), conceptual, to procedural questions (lowest accuracy), where knowledge category is listed in order of cognitive processing required \cite{blooms-taxonomy-2001}. We hypothesize LLMs may not be good at synthesizing multiple pieces of information and formulating new facts.

Second, GPT's overall performance—and its performance within individual security categories—is lower on questions that involve specific products, platforms, or companies (398 out of 900 questions). When a question is related to a product, platform, or company, the probability of the GPT response being fully accurate, thorough, and relevant is 44\%. This probability increases to 60\% for questions that are not associated with such entities.

Lastly, GPT performs particularly poorly in the authentication and antivirus/anti-malware security categories. This underperformance may be attributed to the high proportion (81.15\%) of product-, platform-, or company-specific questions within these categories.

Following this evaluation, we present the most common error and deficiency patterns identified across all three LLMs.  %

\paragraph{Error patterns in LLMs.} Through our manual expert evaluation and inductive thematic analysis, we identified several recurring error patterns across responses from multiple LLMs that have the potential to negatively impact user security. We organized these errors into two high-level categories: content-related errors, which involve incorrect or deficient information, and communication-related errors, which pertain to how the response is conveyed to the user.

Within each high-level category, we further grouped fine-grained error types into mid-level themes that capture broader underlying issues. For example, the mid-level category \textit{\LackAdherenceResearch} encompasses specific patterns such as “Password Guidance,” “HTTPS Over-Reliance,” “App Permission Risks,” and “Lacking Explanations.” In our analysis, we detail each of these patterns and offer hypotheses regarding the potential causes behind these mid-level error groups.

\subsection{Finding: LLM Content Error Patterns}
\label{sec:content_errors}

\paragraph{1) \LackAdherenceResearch.} A key error pattern observed across all three LLMs is a failure to incorporate recent research findings, resulting in responses that are outdated, incorrect, or incomplete. This issue was especially prevalent in areas such as password security (65 responses) and HTTPS reliance (15 responses).

    \noindent \textit{\textbf{Password guidance}:} All evaluated models continued to recommend outdated practices, such as enforcing password complexity and requiring frequent password changes—advice that contradicts current security research and official guidelines. For example, in response to the question “What makes for a secure password?”, GPT advised:
            
            \begin{quote}
                \underline{Complexity}: Include a mix of uppercase letters, lowercase letters, numbers, and special characters (e.g., \!!, @, \#, \$, \%).
            \end{quote}
            \begin{quote}    
                \underline{Regular Updates}: Regularly updating your passwords can help reduce the risk of unauthorized access, especially if there's a chance that your password may have been exposed.
            \end{quote}

    These recommendations are inconsistent with recent research showing that password complexity requirements reduce usability without substantially enhancing security \cite{shay-2016-password-policy}, and that routine password changes—unless prompted by a known breach—offer limited benefit \cite{zhang-2010-password-expiration, chiasson-2015-password-expiration, habib-2018-password-expiration}. Current NIST guidelines also no longer support these practices \cite{grassi-2017-nist-password-guidelines}. This pattern indicates that LLMs are not well-aligned with current research guidance, particularly in security contexts where precision and relevance are essential.
        
    \noindent \textit{\textbf{HTTPS overreliance:}} Another common error across all three LLMs is the over-reliance on HTTPS as an indicator of website legitimacy and safety. The models frequently advised users to trust websites solely based on the presence of HTTPS and a lock icon, without clarifying the protocol’s limitations. For example, Llama suggested that users verify websites by checking for HTTPS, implying it guarantees security and legitimacy. In response to “How can I avoid phishing?”, Llama stated:
        \begin{quote}
            \underline{Check for HTTPS and a lock icon}: When visiting a website, make sure it starts with “https" and has a lock icon in the address bar. \underline{This indicates that the site is secure and legitimate}.
        \end{quote}

    This guidance is misleading. While HTTPS encrypts data in transit, it does not ensure that a website is trustworthy or free from malicious content. According to the 2019 Anti-Phishing Working Group report, 58\% of phishing URLs used HTTPS \cite{apwg2019-HTTPS-phishing-report}, and attackers often exploit this trust to increase the success of phishing attempts \cite{kim-https-phishing-CA-2021}. Additionally, they often omit key points about malicious websites that use HTTPS and the protocol’s broader privacy limitations, reinforcing misconceptions that may endanger users.
    
    By failing to explain that HTTPS guarantees only encryption—not authenticity or safety—LLM responses risk creating a false sense of security. 

    \noindent \textit{\textbf{App permissions risks:}} All three LLMs failed to mention the security and privacy risks associated with app permissions when asked about their necessity. Prior research has identified several concerns, including over-privileged applications \cite{felt-permissions-2011}, deceptive practices by free apps to obtain permissions \cite{chia-permission-risks-2012}, and the misuse of automatically granted permissions without clear disclosure \cite{calciati-auto-permissions-2020}. These risks are particularly important to highlight, as app stores such as Google Play and Apple’s App Store display only broad permission categories (e.g., “Personal info,” “Location”), while more specific details (e.g., access to email or precise location) are hidden deeper in the interface.

    \noindent \textit{\textbf{Lacking explanations:}} Across all three LLMs, responses to questions such as “Why do I need antivirus software?” and “What can happen if someone breaks into my home Wi-Fi?” lacked important research-backed explanations. For the former, LLMs often recommended antivirus software without qualification, despite research showing it is not always necessary or effective in all scenarios \cite{dekoven-security-practices-2019}. For the latter, none of the responses mentioned significant threats posed by local network attackers, such as household fingerprinting and tracking the behavior of occupants \cite{girish-local-iot-2023}.

    We hypothesize that, beyond known LLM limitations, this error may stem from data bias in the training corpus—specifically, an overrepresentation of marketing materials compared to up-to-date security research, which is often paywalled in academic papers and books. This speculation is supported by researchers' observations of similarities between LLM responses and language found on product websites. However, this cannot be confirmed due to the lack of transparency around LLM training datasets.

    \paragraph{2) \Overprescription.} LLMs often portrayed security solutions in an overly optimistic manner, either omitting their limitations or recommending them for inappropriate use cases, resulting in responses that were inaccurate or incomplete. This pattern was observed in 19 responses across all three LLMs.

    \noindent \textit{\textbf{Virtual private networks (VPNs):}} GPT and Gemini incorrectly recommended VPNs for purposes beyond their intended use, such as preventing phishing attacks and encrypting Bluetooth traffic. For the question “What are the benefits of using a VPN?" Gemini responded:
                        \begin{quote}
                        Enhanced Online Security: \uline{A good VPN can protect you from malicious websites and phishing attempts.}
                        \end{quote}
            
            All of them asserted VPNs could prevent identity theft. For the question “How can a VPN help protect against identity theft?" GPT claimed:
                        \begin{quote}
                            \underline{Data Breach Protection}: In the event of a data breach, if you’ve transmitted sensitive information while connected to a VPN, the encrypted data will be less useful to cybercriminals who may have obtained it.
                        \end{quote}

            Additionally, all three models advised using VPNs to secure IoT devices—guidance that is impractical for typical end users. Llama, for example, recommended using a VPN to protect voice assistant data on public networks.
            
            Crucially, none of the models acknowledged that VPNs do not protect against threats from local network adversaries (e.g., on public Wi-Fi) or inherently malicious websites. These omissions contribute to misconceptions about the capabilities and appropriate use of VPNs.

    \noindent \textit{\textbf{Multi-factor authentication (MFA):}} All three LLMs incorrectly recommended using MFA as a defense against spoofing attacks, where attackers impersonate another individual. However, MFA does not prevent users from receiving or engaging with calls from spoofed phone numbers, making this recommendation misleading. For example, in response to the question “What is spoofing?”, Llama advised:
            \begin{quote}
            To protect yourself from spoofing attacks: \uline{Use strong passwords and enable two-factor authentication whenever possible.}
            \end{quote}

    We hypothesize that these error patterns may stem from data quality issues, specifically the presence of unverified claims from marketing materials and blog posts about technology and security products in the training data.

    \paragraph{3) \TransparencyArguments.} LLMs frequently failed to mention key transparency-related factors—such as open-source codebases, public vulnerability disclosure programs, and the historical handling of data breaches—when discussing password managers, VPNs, and MFA providers. These elements are critical for building user trust in security products.

    For example, GPT consistently omitted transparency considerations when recommending password managers, even in response to questions specifically for closed-source options. Gemini occasionally mentioned transparency, while Llama incorrectly claimed that the Google Password Manager Chrome extension is open source, stating:
    \begin{quote}
        \uline{Open-Source Code}: The Google Password Manager Chrome extension is open-source, which allows developers and security experts to review and audit the code for vulnerabilities.
    \end{quote}
    
    In discussions of products with a known history of security incidents—such as LastPass and Okta \cite{lastpass_security_incident-2022-12, lastpass_incident_data-2023-03, lastpass_security_notice_2015, okta_verkada_attack_2021, okta_support_system_unauthorized_access_2023}—GPT did not reference prior breaches, and Llama omitted this information entirely. Gemini only acknowledged such breaches when explicitly asked and did not suggest considering incident history as a criterion for choosing a password manager.
    
    Similarly, none of the models recommended evaluating VPN providers based on their past handling of security incidents, despite documented cases of providers failing to disclose breaches transparently \cite{nordvpn_datacenter_breach_2019}. All three models also neglected to suggest assessing the presence of vulnerability disclosure programs when selecting security tools.
    
    This recurring omission reflects a broader issue with LLM reasoning and may stem, at least in part, from the underrepresentation of content in the training data that explains the importance of transparency in fostering user trust.
    
    \paragraph{4) \MissingThreatAngles.} Across several security-related topics, LLMs failed to address critical threat vectors and appropriate mitigation strategies, potentially leaving users vulnerable. This pattern was observed in 34 responses across the models.

        \noindent \textit{\textbf{Password manager:}} When asked about master password compromise or unusual activity, LLMs consistently omitted key recommendations such as revoking access to the affected account and changing all passwords stored in the manager. These omissions are significant, as failure to revoke access allows attackers continued access, and not updating compromised passwords increases the risk of further breaches.
    
        \noindent \textit{\textbf{IoT and smart devices:}} In response to “How can I make sure my IoT devices are secure?”, none of the models suggested securing devices with WPA2/WPA3 protocols. Additionally, they failed to mention known vulnerabilities such as Blueborne \cite{cert_vu240311}—which enables remote code execution—for questions about Bluetooth security, and did not identify NFC cloning as a threat when asked about NFC risks.
    
        \noindent \textit{\textbf{Antivirus:}} When asked “How can I remove malware from my computer?”, all models incorrectly recommended creating backups prior to malware removal, despite the risk of copying infected files. For the question “What can happen if I use end-of-life antivirus applications?”, GPT failed to mention that such products may contain unpatched vulnerabilities. %
    
        We hypothesize that these omissions result from limitations in both LLM reasoning capabilities and training data coverage, despite the fact that many of these answers are publicly available.

    \paragraph{5) \UISteps.} When responding to procedural or “how-to" questions, all three LLMs frequently provided overly specific, fabricated steps that were either inaccurate or incomplete. In many cases, they omitted steps for relevant platforms or included steps for non-applicable ones. This pattern was the most prevalent among our observations, occurring 169 times across the models.

    For instance, in response to “How do I configure the security features in Safari?”, Gemini suggested:

    \begin{quote}
        General Security Settings: $\Rightarrow$
        Open Safari: Click the Safari icon in your Dock or Applications folder. $\Rightarrow$
        Go to Preferences: Click “Safari" in the menu bar, then select “Preferences". $\Rightarrow$
        Choose “Security": Click the “Security" tab.
    \end{quote}

    These instructions are inaccurate, as Safari no longer includes a “Preferences” option in its menu, rendering the guidance unhelpful—especially for non-technical users seeking to securely configure their browser. %
    
    We hypothesize that this error pattern stems from two main factors: concept drift and hallucination. User interfaces for browsers and security tools change frequently, and documentation may not always reflect these updates promptly, contributing to concept drift. Additionally, many products have similar but not identical features or workflows (e.g., account recovery procedures), and LLMs—due to their probabilistic nature—often generalize or conflate these nuances, leading to inaccurate responses.

    \paragraph{6) \Hallucinations.} Hallucination—a known issue in LLMs—refers to the generation of nonsensical or incorrect information, which can lead to security misinformation, particularly for non-technical users. All three models frequently generated factually incorrect responses, or "hallucinations," that mislead users and compromise information reliability. GPT, for instance, incorrectly linked Bluetooth use on public Wi-Fi to security risks and inaccurately recommended VPNs to secure Bluetooth communications. It also misrepresented device compatibility with antivirus software, suggesting non-existent capabilities across platforms like IoT devices, NAS systems, and smartwatches.

    Llama repeatedly claimed that Google Password Manager uses zero-knowledge proofs—a cryptographic concept not applicable to its actual implementation, which uses zero-knowledge encryption—thus perpetuating a misconception.
    
    Gemini hallucinated that “AVG AntiVirus Hack Alerts” was a scam and incorrectly asserted that “AVG Mobile Security for iOS” does not exist, misleading users by either raising false alarms or downplaying legitimate security tools. These errors likely stem from training data quality issues and inherent reasoning limitations, particularly the prevalence of incorrect information in online sources.

    We hypothesize this error pattern may also be due to data quality and LLM reasoning challenges. Many of the presented facts are well-represented in (incorrect) internet content.

\subsection{Finding: LLM Communication Error Patterns}
\label{sec:communication_errors}

    LLMs use alignment to ensure that content generated aligns with human preferences and values, enhancing usability, safety, and reliability. The following patterns illustrate limitations in current LLM alignment strategies.

    \paragraph{1) \MisinterpretQuestions.} All three models demonstrated difficulty interpreting questions involving homonyms or specific user requirements. For example, when asked how to securely install an IoT camera, GPT provided only physical installation steps, omitting security considerations. In other cases, terms like “viruses," “cookies," “patches," and “secrets" were interpreted incorrectly—yielding biological, culinary, or linguistic definitions rather than relevant technical explanations.

    \paragraph{2) \RestrictiveGuardrails.} Gemini and, to a lesser extent, Llama occasionally refused to answer straightforward user questions, citing safety concerns. This behavior was most prominent in Gemini, which declined to respond to a question such as “What are the five most common passwords nowadays?” Llama exhibited this only once. In response to “What is an example of a secure password?”, Gemini replied: \begin{footnotesize}\textit{“It's great you're thinking about password security! I can't give you a specific example of a secure password, because that would be a security risk itself."}
    \end{footnotesize}

    Additionally, questions like “Can a password be attacked by brute force?”, “How can hackers use public Wi-Fi to steal my information?”, and “How do I scan my PC for potential threats with AVG AntiVirus?” were blocked by Gemini’s default safety configurations until adjustments were made. While safety measures are important, achieving a balance is crucial to maintain the models’ usability.
    
    \paragraph{3) \IndirectResponses.} Users tend to have short attention spans when reading online \cite{Schade2018, Nielsen1996, Nielsen1997}, making it important for responses to be direct and relevant. However, LLMs frequently begin with information already implied by the question or with generic, encouraging, or patronizing statements—such as “That’s a great question!” or “It’s important to feel secure about your passwords.” We refer to this indirect style as LLM-splaining. All three models exhibited this behavior frequently, with 260 recorded instances. For example, in response to “Can I trust Google Password Manager?”, GPT began with: \begin{footnotesize}\textit{“Google Password Manager is a feature offered by Google that stores and manages passwords for various websites and applications."}\end{footnotesize}
    
    \paragraph{4) \ResponsesGearedTowardsDevelopers.} At times, the models provided responses that were too technical for general end-users, instead targeting developers or system administrators. For example, in response to “How can phishing be prevented?”, Llama recommended setting up SPF, DKIM, and DMARC—solutions more appropriate for IT professionals. Similarly, when asked about the importance of HTTPS, models emphasized SEO and reputation impact, which are concerns specific to web developers. Responses to “How do I install Okta?” were for system administrators. 
     
     We hypothesize, these misalignments may stem from rigid safety constraints and an overrepresentation of developer-focused documentation in the training data.

\section{Guidance and Open Problems}
\label{sec:discussion}

Previous research has found that decision making in AI-assisted contexts improves when humans and AI complement each other, taking their respective weaknesses into account \cite{bansal2019beyond}. Without an understanding of AI weaknesses, humans may over- or under-rely on AI for assistance \cite{schoeffer2024ai}. On the human side, three well-established weaknesses or challenges for which AI can \textit{potentially} compensate are: 1. The challenge of keeping up with security guidance and tools, 2. User mental models of security that may not be aligned with good security practices \cite{wash2011influencing} and 3. Barriers to user acceptance of security advice \cite{redmiles2016learned}.  We consider the findings of~\Cref{findings}  in the context of these challenges to distill guidance for end-users and model developers toward improving human-LLM interaction in the user security assistance context. We also identify places where additional research is needed to extend this guidance.

\subsection{User Guidance}

We discuss what our evaluation suggests regarding security topics and question types best suited to LLMs, how user questions are expressed (i.e., prompt engineering), and how users can assess LLM responses. Note that while we report performance numbers for GPT only, all the errors discussed in this section were also evident with Llama and Gemini, except where otherwise indicated.\footnote{Our analysis of Llama and Gemini responses focused on questions that GPT responded to incorrectly, so while we know similar errors occur with Llama and Gemini, we do not know how they performed overall.}

\paragraph{When to use LLMs.} Based on our evaluation, we discourage users from relying on LLMs for procedural questions (e.g., ``How do I…?’’), regardless of security topic. Our evaluation found a high rate of errors (31\% for GPT) and omissions of important information (43\% for GPT) in response to procedural questions.

Compatible with research in other domains, LLMs struggled with questions requiring reasoning or comparison, such as security trends. For example, when asked ``What age groups most report fraud or scams?'', GPT calls out 4 age groups, including teens, adults and seniors, when in fact seniors are less likely to report fraud and scams than younger people \cite{ftc_scams_all_ages_2022}.

Another challenging area is questions that are sensitive to recent events, such as queries regarding trust and reputation. For example, when asked ``Can I get a free password manager?'', GPT suggests Lastpass (among others), a password manager that is no longer recommended due to recent breaches.

Finally, we encourage care around questions in the areas of authentication and antivirus, particularly when asking product, platform, or company-related questions, since significant weakness in response quality was found in both. GPT errors exceeded 30\% in both categories, and the same weakness was present in both Llama and Gemini responses. 

Areas of strength for LLMs are definitions of security concepts (e.g., ``What is malware?''), basic risk assessment questions (e.g., ``Will 2 factor authentication protect me?'') and core functionality of established products (e.g., ``How does Keeper password manager work?''). This is compatible with the fact that factual and conceptual questions are the most likely to be answered correctly (\Cref{tab:gpt-evaluation-dist-by-topics}). 

\paragraph{How to communicate with LLMs.} We used questions from published FAQs and question templates that our review found to be written in ``plain’’ English (i.e., no technical, rare or otherwise lesser-known words and phrases \cite{williams2004legal}). We conducted a small experiment to test the sensitivity of all three LLMs to rephrased questions---which are organic user queries seen on Google search---from this corpus (Section~\ref{sec:reproducibility}) and did not find sensitivity to phrasing.

In addition, several other open prompt engineering questions would allow for important extensions of user guidance in this area. For example, we found a substantial number of indirect responses (17\% with GPT) that delay information delivery and could cause user attention to shift elsewhere. We term a portion of these responses ``LLM-splaining’’ because they provide information that is likely already known by the user. The unnecessary repetition of information may diminish user trust \cite{zeffane2011communication}. For example, to the question, ``How do I set up Okta Verify?’’, GPT responds with 2 sentences explaining Okta Verify before beginning to answer the user question: ``Okta Verify is an authentication app that works in conjunction with Okta, a secure identity management service. It provides a secure way…’’. Prompts that encourage LLMs to consistently address user questions directly are an open problem.

Finally, we note that while our questions did not contain any sensitive information, security questions can involve sensitive data so user prompts may be sensitive, particularly in light of patterns of over-sharing with LLMs that have been found \cite{zhang2024s}. The relationship between LLM response quality and the inclusion of sensitive information is largely unexplored and would be a useful addition to user guidance.

\paragraph{User assessment of LLM responses.} While this may change with LLM-enabled search, none of the LLMs we experimented with routinely cite sources when answering end-user security questions. Indeed, URLs are present in less than 5\% of the GPT answers in our evaluation. We recommend users explicitly ask for authoritative sources supporting LLM responses, and consider requesting that LLMs find supporting data from government or nonprofit sources of the user’s choosing (e.g., NIST, FTC).

As discussed in \cite{barman2024beyond}, ``simple heuristics to detect potential
errors or misinformation from LLMs’’ should be part of a user toolkit for safe interaction with LLMs. Towards such a toolkit we see evidence of two heuristics in our data given that errors in some of the evaluation criteria are predictive of other errors. In particular, indirect GPT responses ($n=153$) have errors in accuracy, thoroughness and relevance with probability greater than $.58$. Similarly, irrelevant GPT answers ($n=15$) have errors in accuracy, thoroughness or directness with probably greater than $.73$. The support for each of these heuristics is relatively low, so it is an open problem to confirm that these heuristics hold. That said, they suggest a promising direction for helping users safely interact with LLMs since indirect responses and irrelevant content may be more easily recognized than the other criteria and serve as an early user warning that a response may be low quality. While there has been recent progress in computational approaches to predicting hallucinations \cite{farquhar2024detecting}, \textit{user-friendly} heuristics are an important open problem.

\add{We note that our study %
informs user guidance for interacting with LLMs but does not explore user perception of LLM responses. Understanding the applicability and understandability of LLM responses to user security questions is an open problem.}

\subsection{Developer Guidance}
In this section, we discuss the primary areas of performance problems in our evaluation, organized according to the user challenges described at the beginning of Section~\ref{sec:discussion}. While we suggest directions for improvements, we are not prescriptive, as model developers are better positioned to identify root causes and craft solutions.

\paragraph{Generating correct security content.} Our evaluation suggests augmenting LLM training data with more security research and reducing the support of security product marketing materials will improve LLM performance. For example, all LLMs in our evaluation routinely provide stale password guidance despite ample research attesting to the usability shortcomings of such guidance. In addition, the privacy and security risks of default app permissions, a topic well-covered by research \cite{felt-permissions-2011, chia-permission-risks-2012, calciati-auto-permissions-2020}, are often overlooked in LLM responses. Finally, the incorrectly inflated capabilities of certain security technologies (e.g., HTTPS) suggest marketing materials may have high support in LLM training data. For example, “Authy takes security very seriously", and “Dashlane is a well-established and reputable password manager that has earned the trust of millions of users worldwide." Overly favorable product content may be perceived as marketing information and can be devalued by users \cite{redmiles2016learned}.

Questions involving specific companies, products, or platforms are often challenging for LLMs. For example, when no company, product, or platform is mentioned, more than 60\% of GPT responses are perfectly accurate, thorough, and relevant. In contrast, this number drops to less than 44\% when a specific company, product, or platform is part of the prompt. We suggest that the presence of a company name, product, or platform may be useful for developers in determining when to prioritize authoritative sources.

Finally, while few in number, our evaluation did surface 10 examples in which the human-LLM interaction completely failed because the LLM was not responsive to the question. In 7 of these (6 for Gemini and 1 for Llama), this appeared to be due to mistriggered guardrails; the 3 others were GPT misinterpretations (e.g., to the question, ``Can you write your login credentials down, really?’’ GPT responds, ``I'm sorry for any confusion, but as an AI, I don't possess personal login credentials…”). We suggest that developers consider how to better confirm the user intent before delivering a non-answer.

In short, while those with a complete view of the model development process are best positioned to determine how to address these errors as well as the hallucinations and omitted risks discussed in Section~\ref{findings}, we emphasize that answers to questions in these areas are well documented in authoritative security sources (e.g., peer-reviewed research papers, government reports, product and platform documentation, etc.) and so it appears models either do not have sufficient access to this data or aren’t prioritizing data sources appropriately.

\paragraph{Aligning mental models with sound security practices.} Our question corpus was gathered to broadly represent user concerns in security, rather than focus on areas known to have potentially problematic user models. That said, we do see evidence of known user mental models in LLM answers. For example, \cite{wash2010folk} found that the term ``hacker'' is often used to broadly describe bad actors in a security context regardless of the actor's attack strategy or relationship to the user. We see a similar broad use of the term in LLM responses, GPT characterizes actors launching technical attacks as hackers (``The longer the password, the more combinations a hacker has to try to crack it.") as well as scammers (``Hackers can use sophisticated phishing schemes to trick users"). In addition, the inflation of the protective power of HTTPS noted in Section~\ref{findings} was also found in \cite{herbert2023world} among participants in India.

We emphasize that LLM perpetuation of user mental models is only problematic if those models are associated with risky security practices. For example, ``hacker'' terminology may suppress the fact that security risks also exist with people a user knows (e.g., intimate partner or elder abuse).

Finally, we note while some LLM error patterns may not specifically perpetuate problematic mental models, they may still contribute to areas of confusion. While we are unaware of previous research around the mental models of linking VPNs to phishing protection, the protective power of VPNs is a well-established area of user confusion (e.g., \cite{dutkowska2022and, binkhorst2022security}).

We encourage model developers to use security user mental models research to explore potential areas of bias and consider relying more heavily on authoritative sources in those areas.
 
\paragraph{Encouraging adoption of security information.} We recommend developers tailor the balance between content and sources based on the nature of the user question. In particular, consider deferring to sources in the cases of procedural and UI-dependent questions,  or when contact information or URLs are part of the response. For instance, GPT provided email addresses in 6 responses, none of which were valid at the time of our evaluation. Also, in 3 responses, GPT provided 5 (out of 123 URLs across 47 responses) inaccessible URLs. Our evaluation indicates making strategic choices between the balance of content and sources in LLM answers will improve response quality and may increase user acceptance; there is evidence that acceptance can be sensitive to the authoritativeness of the source, for example in areas that are less familiar to the user \cite{redmiles2016learned}. 

Given that user security questions may be time-sensitive, we recommend LLMs use a direct communication style in response to such questions. In addition, developers should consider signaling to the user when they are not able to deliver a high-confidence response, similar to the warnings many LLMs already provide about potential limitations due to the date range of training data. Finally, we raise the question of whether explanations can also help the end user better detect errors in the end user security context \cite{bansal2021does}.

\section{Limitations and Future Work} 
\subsection{Representation of user queries in our corpus}
\label{sec:represent}

The findings of this study depend on the user security questions used for the assessment. We therefore acknowledge that our assessment is closely tied to our efforts to curate end-user security questions.

To validate the real-world relevance of our question corpus, we compared it against user-generated security questions from Reddit and Quora. For each of the 900 study questions, we retrieved up to 20 top Google search results using site-specific queries (“site:reddit.com” and “site:quora.com”). We extracted user questions either directly from the URL (Quora) or via API (Reddit), and computed their cosine similarity with the corresponding study question. Our analysis showed that over 71\% of study questions had a corresponding Reddit or Quora question with a similarity score of at least 0.75, and more than 22\% had a similarity of at least 0.9, indicating strong alignment with real-world user concerns.

As part of our analysis to determine reliability for paraphrased questions (\Cref{appendix-prompt-sensitivity}), we showed that our findings are equivalent to the use of questions in the voice of users and would not change with their inclusion. In addition, questions are listed in the FAQs of consumer-focused organizations because they are \textit{frequently} asked by users; presumably, they are \textit{representative} of broadly held user concerns and less characterized by query or interaction styles that are known to change over time \cite{evolution-of-search}, so our study would be relevant over time by not including the questions in the users' voices. %

Still, since we have not recruited users for our study, we acknowledge that our questions might miss the exact natural language users use to interact with LLMs. %
Users' natural language of interaction with generative AI, exhaustive assessment of security-related questions \add{with prompt engineering}, \add{and exploration of user perception of LLM responses} are left as future work.

As with any qualitative study, our manual evaluation of LLM responses is subject to potential human error and interpretation bias. To mitigate this, we conducted a four-way evaluation on a subset of 72 responses, achieving an IRR categorized as “substantial.” Additionally, all individually rated responses underwent dual verification, and any disagreements were resolved through discussion prior to inclusion in the final results. 

\subsection{Future Work}
Another important aspect of digital safety is privacy; the assessment of LLMs in the privacy domain is left as future work.%

The error reasons are hypotheses based on previously published LLM research. Some of these hypotheses are challenging to verify, in particular, training data-related hypotheses depend on access to model training data. Also, our recommendations for developers are based on prior research, but we have not verified that the recommendations solve the identified issues when followed. %
Our usability recommendations have not yet been tested on actual users to measure their efficacy. %

Researchers should evaluate AI-powered search engines that generate source-based responses (e.g., ChatGPT search). During our study, these tools were newly introduced and lacked accessible APIs, limiting large-scale evaluation. Future research should examine the sources these engines suggest and how they are used to construct responses.

Our findings are limited to the specific LLMs evaluated in this study and may not generalize to other models or future versions of the same models, particularly as they continue to evolve. Developing an automated evaluation system that can assess open-ended LLM responses with expert-level accuracy represents a promising direction for future research. Our expert-labeled dataset of question–response pairs could serve as a valuable resource for training or benchmarking such systems. Additionally, establishing a benchmark for automated evaluation of LLM performance on open-ended security-related questions would enable systematic, quantitative comparisons across models and versions over time. We leave these avenues for future work.
\if \conference1
\section{Ethics Considerations}
\label{sec:compliance}
\else
\section{Data and Ethical Considerations}
\fi

Our study has been reviewed from an ethical standpoint and determined to be out of scope of the authors' university IRB approval process.
Our study is approved by the legal and ethical review process  of the industry author's organization.

While we have evaluated LLM responses to questions that reflect user concerns, none of the questions include personal user information and, to the best of our knowledge, none are verbatim user questions and so are unlikely to include identifying language. In addition, we used paid subscriptions to the GPT and Gemini APIs which ensures user data (prompts) and responses are not used for future model development, as per OpenAI and Google privacy policies at the time of our experiments.

Finally, we note that while we surfaced error patterns in LLM responses that negatively impact the user experience, we do not believe any present an urgent risk, hence we are not planning to disclose our findings to model development companies prior to publication.

\if \conference1
\section{Open Science}
\label{sec:open-science}
Our study evaluates security information provided by LLMs on 7 security topics. We will make the question corpus and our codebooks publicly available to support future research. We will also make expert-rated LLM responses available to researchers upon request. 
\fi

\section*{Acknowledgement}
\noindent This work was partially supported by NSF award CNS-2232655. Vijay Prakash's contributions are supported in part by a research award from JPMorgan Chase.

This paper was prepared for informational purposes in part by the Chief Data \& Analytics Office of JPMorgan Chase \& Co. and its affiliates (``J.P. Morgan'') and is not a product of the Research Department of J.P. Morgan. J.P. Morgan makes no representation and warranty whatsoever and disclaims all liability, for the completeness, accuracy or reliability of the information contained herein.  This document is not intended as investment research or investment advice, or a recommendation, offer or solicitation for the purchase or sale of any security, financial instrument, financial product or service, or to be used in any way for evaluating the merits of participating in any transaction, and shall not constitute a solicitation under any jurisdiction or to any person, if such solicitation under such jurisdiction or to such person would be unlawful.
\printbibliography

@article{sharma2023towards,
  title={Towards understanding sycophancy in language models},
  author={Sharma, Mrinank and Tong, Meg and Korbak, Tomasz and Duvenaud, David and Askell, Amanda and Bowman, Samuel R and Cheng, Newton and Durmus, Esin and Hatfield-Dodds, Zac and Johnston, Scott R and others},
  journal={arXiv preprint arXiv:2310.13548},
  year={2023}
}

@ARTICLE{mchugh2012-irr,
  title     = "Interrater reliability: the kappa statistic",
  author    = "McHugh, Mary L",
  journal   = "Biochem. Med. (Zagreb)",
  publisher = "Croatian Society for Medical Biochemistry and Laboratory
               Medicine",
  volume    =  22,
  number    =  3,
  pages     = "276--282",
  year      =  2012,
  language  = "en"
}

@misc{apwg2019-HTTPS-phishing-report,
  author       = {APWG (Anti-Phishing Working Group)},
  title        = {Phishing Activity Trends Report, 1st Quarter 2019},
  year         = {2019},
  url          = {https://docs.apwg.org/reports/apwg_trends_report_q1_2019.pdf},
  note         = {Accessed: 2024-08-08}
}

@misc{MetaLlama3launchblog,
  author = {{Meta AI}},
  title = {{Meta Llama 3 Release Blogpost}},
  howpublished = {\url{https://ai.meta.com/blog/meta-Llama-3/}},
  year = {2024},
  note = {Accessed: 2024-05-12}
}

@misc{Llama3git,
  author = {{Meta AI}},
  title = {{Llama 3}},
  howpublished = {\url{https://github.com/meta-Llama/Llama3}},
  year = {2024},
  note = {Accessed: 2024-05-12}
}

@misc{inan2023Llama,
      title={Llama Guard: LLM-based Input-Output Safeguard for Human-AI Conversations}, 
      author={Hakan Inan and Kartikeya Upasani and Jianfeng Chi and Rashi Rungta and Krithika Iyer and Yuning Mao and Michael Tontchev and Qing Hu and Brian Fuller and Davide Testuggine and Madian Khabsa},
      year={2023},
      eprint={2312.06674},
      archivePrefix={arXiv},
      primaryClass={cs.CL}
}

@misc{bhatt2023purple,
      title={Purple Llama CyberSecEval: A Secure Coding Benchmark for Language Models}, 
      author={Manish Bhatt and Sahana Chennabasappa and Cyrus Nikolaidis and Shengye Wan and Ivan Evtimov and Dominik Gabi and Daniel Song and Faizan Ahmad and Cornelius Aschermann and Lorenzo Fontana and Sasha Frolov and Ravi Prakash Giri and Dhaval Kapil and Yiannis Kozyrakis and David LeBlanc and James Milazzo and Aleksandar Straumann and Gabriel Synnaeve and Varun Vontimitta and Spencer Whitman and Joshua Saxe},
      year={2023},
      eprint={2312.04724},
      archivePrefix={arXiv},
      primaryClass={cs.CR}
}

@misc{PurpleLlama,
  author = {{Meta AI}},
  title = {{Purple Llama}},
  howpublished = {\url{https://github.com/meta-Llama/PurpleLlama/tree/main}},
  year = {2024},
  note = {Accessed: 2024-05-12}
}

@misc{Gemini,
  author = {{Google}},
  title = {{Gemini}},
  howpublished = {\url{https://gemini.google.com/app}},
  year = {2024},
  note = {Accessed: 2024-05-12}
}

@misc{GeminiReleaseBlog,
  author = {{Google}},
  title = {{Gemini Release Blogpost}},
  howpublished = {\url{https://blog.google/technology/ai/google-gemini-ai/}},
  year = {2024},
  note = {Accessed: 2024-05-12}
}

@misc{geminiteam2024geminiv1.5,
      title={Gemini 1.5: Unlocking multimodal understanding across millions of tokens of context}, 
      author={Gemini Team and Machel Reid and Nikolay Savinov and Denis Teplyashin and Dmitry and Lepikhin et al.},
      year={2024},
      eprint={2403.05530},
      archivePrefix={arXiv},
      primaryClass={cs.CL}
}

@misc{geminiteam2024geminiv1,
      title={Gemini: A Family of Highly Capable Multimodal Models}, 
      author={Gemini Team and Rohan Anil and Sebastian Borgeaud and Jean-Baptiste Alayrac and Jiahui Yu and Radu Soricut et al.},
      year={2024},
      eprint={2312.11805},
      archivePrefix={arXiv},
      primaryClass={cs.CL}
}

@misc{GPT4Research,
  author = {{OpenAI}},
  title = {{GPT-4 Release Blogpost}},
  howpublished = {\url{https://openai.com/index/gpt-4-research/}},
  year = {2024},
  note = {Accessed: 2024-05-12}
}

@misc{GPT4SystemCard,
  author = {{OpenAI}},
  title = {{GPT-4 Model Card}},
  howpublished = {\url{https://cdn.openai.com/papers/gpt-4-system-card.pdf}},
  year = {2024},
  note = {Accessed: 2024-05-12}
}

@misc{openaistories,
  author = {{OpenAI}},
  title = {{ChatGPT}},
  howpublished = {\url{https://openai.com/news/stories/}},
  year = {2024},
  note = {Accessed: 2024-05-12}
}

@misc{openai2024gpt4,
      title={GPT-4 Technical Report}, 
      author={OpenAI and Josh Achiam and Steven Adler and Sandhini Agarwal et al.},
      year={2024},
      eprint={2303.08774},
      archivePrefix={arXiv},
      primaryClass={cs.CL}
}

@misc{NIST,
  author = {{National Institute of Standards and Technology}},
  title = {{NIST}},
  howpublished = {\url{https://www.nist.gov/}},
  note = {Accessed: May 11, 2024},
  year = {2024}
}

@misc{NCSC,
  author = {{National Cyber Security Centre}},
  title = {{NCSC}},
  howpublished = {\url{https://www.ncsc.gov.uk/section/advice-guidance/all-topics}},
  note = {Accessed: May 11, 2024},
  year = {2024}
}

@misc{CISA,
  author = {{Cybersecurity and Infrastructure Security Agency}},
  title = {{CISA}},
  howpublished = {\url{https://www.cisa.gov/}},
  note = {Accessed: May 11, 2024},
  year = {2024}
}

@misc{NCA,
  author = {National Cybersecurity Alliance},
  title  = {{NCA}},
  howpublished = {\url{https://staysafeonline.org/}},
  note = {Accessed: May 11, 2024},
  year = {2024}
}

@misc{evolution-of-search,
  title = {The evolution of Search},
  howpublished = {\url{https://www.thinkwithgoogle.com/intl/en-apac/consumer-insights/consumer-trends/evolution-search/}},
  note = {Accessed: Aug 29, 2023},
  year = {2023}
}

@misc{pcmag-chatGPT-v-Google,
  title = {When Will ChatGPT Replace Search? Maybe Sooner Than You Think},
  howpublished = {\url{https://www.pcmag.com/news/when-will-chatgpt-replace-search-engines-maybe-sooner-than-you-think}},
  note = {Accessed: Aug 29, 2023},
  year = {2023}
}

@misc{google-query-rec,
  title = {Google Search Related Questions API},
  howpublished = {\url{https://github.com/lagranges/people_also_ask}},
  note = {Accessed: Aug 29, 2023},
  year = {2022}
}

@misc{google-auto-complete-api,
  title = {Google auto-complete API},
  howpublished = {\url{https://stackoverflow.com/a/13623267/2229448}},
  note = {Accessed: Aug 29, 2023},
  year = {2012}
}

@misc{tensorflow-USE,
  title = {TenserFlow Universal Sentence Encoder},
  howpublished = {\url{https://www.tensorflow.org/hub/tutorials/semantic_similarity_with_tf_hub_universal_encoder}},
  note = {Accessed: Aug 29, 2023},
  year = {2023}
}

@misc{lastpass_incident_data-2023-03,
  author       = {LastPass},
  title        = {Incident Data Report},
  year         = {2023},
  howpublished = {\url{https://support.lastpass.com/s/document-item?bundleId=lastpass&topicId=LastPass/incident-data.html}},
  note         = {Accessed: 2024-08-08}
}

@misc{lastpass_security_incident-2022-12,
  author       = {LastPass},
  title        = {Notice of Security Incident},
  year         = {2022},
  howpublished  = {\url{https://blog.lastpass.com/posts/2022/12/notice-of-security-incident}},
  note         = {Accessed: 2024-08-08}
}

@misc{lastpass_security_notice_2015,
  author       = {LastPass},
  title        = {LastPass Security Notice},
  year         = {2015},
  howpublished  = {\url{https://blog.lastpass.com/posts/2015/06/lastpass-security-notice}},
  note         = {Accessed: 2024-08-08}
}

@misc{okta_verkada_attack_2021,
  author       = {Okta},
  title        = {CISO's Perspective: Recent Verkada Cyber Attack},
  year         = {2021},
  howpublished = {\url{https://sec.okta.com/articles/2021/03/csos-perspective-recent-verkada-cyber-attack}},
  note         = {Accessed: 2024-08-08}
}

@misc{okta_support_system_unauthorized_access_2023,
  author       = {Okta},
  title        = {Tracking Unauthorized Access to Okta's Support System},
  year         = {2023},
  howpublished = {\url{https://sec.okta.com/articles/2023/10/tracking-unauthorized-access-oktas-support-system}},
  note         = {Accessed: 2024-08-08}
}

@misc{nordvpn_datacenter_breach_2019,
  author       = {NordVPN},
  title        = {Official Response to Datacenter Breach},
  year         = {2019},
  howpublished = {\url{https://nordvpn.com/blog/official-response-datacenter-breach/}},
  note         = {Accessed: 2024-08-08}
}

@misc{ftc_scams_all_ages_2022,
  author       = {Federal Trade Commission (FTC)},
  title        = {Who Experiences Scams? A Story for All Ages},
  year         = {2022},
  howpublished = {\url{https://www.ftc.gov/news-events/data-visualizations/data-spotlight/2022/12/who-experiences-scams-story-all-ages}},
  note         = {Accessed: 2024-08-08}
}

@misc{ftc_consumer_sentinal_scam_by_age_2023,
  author       = {FTC},
  title        = {Reported Frauds and Losses by Age},
  year         = {2023},
  howpublished = {\url{https://public.tableau.com/shared/BFHTJ9PMC?:display_count=n&:origin=viz_share_link}},
  note         = {Accessed: 2024-08-08}
}

@misc{Schade2018,
  author    = {Amy Schade},
  title     = {Inverted Pyramid: Writing for Comprehension},
  year      = {2018},
  note      = {Accessed: 2024-10-04},
  howpublished = {\url{https://www.nngroup.com/articles/inverted-pyramid/}},
}

@misc{Nielsen1996,
  author    = {Jakob Nielsen},
  title     = {Inverted Pyramids in Cyberspace},
  year      = {1996},
  note      = {Accessed: 2024-10-04},
  howpublished = {\url{https://www.nngroup.com/articles/inverted-pyramids-in-cyberspace/}},
}

@misc{Nielsen1997,
  author    = {Jakob Nielsen},
  title     = {How Users Read on the Web},
  year      = {1997},
  note      = {Accessed: 2024-10-04},
  howpublished = {\url{https://www.nngroup.com/articles/how-users-read-on-the-web/}},
}

@misc{Liu-ai-usage-2024,
  author    = {Yan Liu, He Wang},
  title     = {Who on Earth Is Using Generative AI ?},
  year      = {2024},
  note      = {Accessed: 2024-10-04},
  howpublished = {\url{https://documents1.worldbank.org/curated/en/099720008192430535/pdf/IDU15f321eb5148701472d1a88813ab677be07b0.pdf}},
}

@misc{google-autocomplete-2024,
  author    = {Google},
  title     = {How Google autocomplete predictions work},
  year      = {2025},
  note      = {Accessed: 2025-01-21},
  howpublished = {\url{https://support.google.com/websearch/answer/7368877}},
}

@article{braun-2006-thematic-analysis,
author = {Virginia Braun and Victoria Clarke},
title = {Using thematic analysis in psychology},
journal = {Qualitative Research in Psychology},
volume = {3},
number = {2},
pages = {77--101},
year = {2006},
publisher = {Routledge},
doi = {10.1191/1478088706qp063oa},
URL = {
        https://www.tandfonline.com/doi/abs/10.1191/1478088706qp063oa
}
}

@online{codebook,
  title     = {Codebook},
  year      = {2025},
  note      = {Accessed: 2025-01-23},
  url       = {https://docs.google.com/document/d/e/2PACX-1vQ99tup4hftyJMDTYlRgL41AM560v-fFqWfSQzb2-G1emV-j_etghUB0-DYL0j9cMSQX7AYjF9d5Ojt/pub},
}

@misc{GoogleAuthenticatorSetup,
  title     = {Set up Google Authenticator for your Google Account},
  year      = {2025},
  note      = {Accessed: 2025-01-23},
  howpublished = {\url{https://support.google.com/accounts/answer/1066447}},
}

@misc{FTC2024-public-wifi,
  author    = {Federal Trade Commission},
  title     = {Are Public Wi-Fi Networks Safe? What You Need to Know},
  year      = {2024},
  month     = {02},
  note      = {Accessed: 2025-01-23},
  howpublished = {\url{https://consumer.ftc.gov/articles/are-public-wi-fi-networks-safe-what-you-need-know}},
}

@misc{NSA-wireless,
  author    = {NSA},
  title     = {Securing Wireless Devices in Public Settings},
  year      = {2021},
  note      = {Accessed: 2025-01-23},
  howpublished = {\url{https://media.defense.gov/2021/Jul/29/2002815141/-1/-1/0/CSI_SECURING_WIRELESS_DEVICES_IN_PUBLIC.PDF}},
}

@String{Computing = "Computing" }

@String{Computer = "{IEEE} Computer" }

@String{Academic = "Academic Press" }

@String{Springer = "Springer-Verlag" }

@misc{brown2020language,
      title={Language Models are Few-Shot Learners}, 
      author={Tom B. Brown and Benjamin Mann and Nick Ryder and Melanie Subbiah and Jared Kaplan and Prafulla Dhariwal and Arvind Neelakantan and Pranav Shyam and Girish Sastry and Amanda Askell and Sandhini Agarwal and Ariel Herbert-Voss and Gretchen Krueger and Tom Henighan and Rewon Child and Aditya Ramesh and Daniel M. Ziegler and Jeffrey Wu and Clemens Winter and Christopher Hesse and Mark Chen and Eric Sigler and Mateusz Litwin and Scott Gray and Benjamin Chess and Jack Clark and Christopher Berner and Sam McCandlish and Alec Radford and Ilya Sutskever and Dario Amodei},
      year={2020},
      eprint={2005.14165},
      archivePrefix={arXiv},
      primaryClass={cs.CL}
}

@misc{hendrycks2021measuring,
      title={Measuring Massive Multitask Language Understanding}, 
      author={Dan Hendrycks and Collin Burns and Steven Basart and Andy Zou and Mantas Mazeika and Dawn Song and Jacob Steinhardt},
      year={2021},
      eprint={2009.03300},
      archivePrefix={arXiv},
      primaryClass={cs.CY}
}

@misc{kojima2023large,
      title={Large Language Models are Zero-Shot Reasoners}, 
      author={Takeshi Kojima and Shixiang Shane Gu and Machel Reid and Yutaka Matsuo and Yusuke Iwasawa},
      year={2023},
      eprint={2205.11916},
      archivePrefix={arXiv},
      primaryClass={cs.CL}
}

@inproceedings{rajani-etal-2019-explain,
    title = "Explain Yourself! Leveraging Language Models for Commonsense Reasoning",
    author = "Rajani, Nazneen Fatema  and
      McCann, Bryan  and
      Xiong, Caiming  and
      Socher, Richard",
    editor = "Korhonen, Anna  and
      Traum, David  and
      M{\`a}rquez, Llu{\'\i}s",
    booktitle = "Proceedings of the 57th Annual Meeting of the Association for Computational Linguistics",
    month = jul,
    year = "2019",
    address = "Florence, Italy",
    publisher = "Association for Computational Linguistics",
    url = "https://aclanthology.org/P19-1487",
    doi = "10.18653/v1/P19-1487",
    pages = "4932--4942",
}

@inproceedings{shwartz-etal-2020-unsupervised,
    title = "Unsupervised Commonsense Question Answering with Self-Talk",
    author = "Shwartz, Vered  and
      West, Peter  and
      Le Bras, Ronan  and
      Bhagavatula, Chandra  and
      Choi, Yejin",
    editor = "Webber, Bonnie  and
      Cohn, Trevor  and
      He, Yulan  and
      Liu, Yang",
    booktitle = "Proceedings of the 2020 Conference on Empirical Methods in Natural Language Processing (EMNLP)",
    month = nov,
    year = "2020",
    address = "Online",
    publisher = "Association for Computational Linguistics",
    url = "https://aclanthology.org/2020.emnlp-main.373",
    doi = "10.18653/v1/2020.emnlp-main.373",
    pages = "4615--4629",
}

@article{Lu_2018,
   title={Learning under Concept Drift: A Review},
   ISSN={2326-3865},
   url={http://dx.doi.org/10.1109/TKDE.2018.2876857},
   DOI={10.1109/tkde.2018.2876857},
   journal={IEEE Transactions on Knowledge and Data Engineering},
   publisher={Institute of Electrical and Electronics Engineers (IEEE)},
   author={Lu, Jie and Liu, Anjin and Dong, Fan and Gu, Feng and Gama, Joao and Zhang, Guangquan},
   year={2018},
   pages={1–1} }

@misc{bommasani2022opportunitiesrisksfoundationmodels,
      title={On the Opportunities and Risks of Foundation Models}, 
      author={Rishi Bommasani and Percy Liang et al.},
      year={2022},
      eprint={2108.07258},
      archivePrefix={arXiv},
      primaryClass={cs.LG},
      url={https://arxiv.org/abs/2108.07258}, 
}

@inproceedings{mt-bench-zheng-2023,
 author = {Zheng, Lianmin and Chiang, Wei-Lin and Sheng, Ying and Zhuang, Siyuan and Wu, Zhanghao and Zhuang, Yonghao and Lin, Zi and Li, Zhuohan and Li, Dacheng and Xing, Eric and Zhang, Hao and Gonzalez, Joseph E and Stoica, Ion},
 booktitle = {Advances in Neural Information Processing Systems},
 editor = {A. Oh and T. Naumann and A. Globerson and K. Saenko and M. Hardt and S. Levine},
 pages = {46595--46623},
 publisher = {Curran Associates, Inc.},
 title = {Judging LLM-as-a-Judge with MT-Bench and Chatbot Arena},
 url = {https://proceedings.neurips.cc/paper_files/paper/2023/file/91f18a1287b398d378ef22505bf41832-Paper-Datasets_and_Benchmarks.pdf},
 volume = {36},
 year = {2023}
}

@inproceedings{zhang-etal-2024-safetybench,
    title = "{S}afety{B}ench: Evaluating the Safety of Large Language Models",
    author = "Zhang, Zhexin  and
      Lei, Leqi  and
      Wu, Lindong  and
      Sun, Rui  and
      Huang, Yongkang  and
      Long, Chong  and
      Liu, Xiao  and
      Lei, Xuanyu  and
      Tang, Jie  and
      Huang, Minlie",
    editor = "Ku, Lun-Wei  and
      Martins, Andre  and
      Srikumar, Vivek",
    booktitle = "Proceedings of the 62nd Annual Meeting of the Association for Computational Linguistics (Volume 1: Long Papers)",
    month = aug,
    year = "2024",
    address = "Bangkok, Thailand",
    publisher = "Association for Computational Linguistics",
    url = "https://aclanthology.org/2024.acl-long.830",
    doi = "10.18653/v1/2024.acl-long.830",
    pages = "15537--15553",
}

@article{chang-llms-eval-survey-2024,
author = {Chang, Yupeng and Wang, Xu and Wang, Jindong and Wu, Yuan and Yang, Linyi and Zhu, Kaijie and Chen, Hao and Yi, Xiaoyuan and Wang, Cunxiang and Wang, Yidong and Ye, Wei and Zhang, Yue and Chang, Yi and Yu, Philip S. and Yang, Qiang and Xie, Xing},
title = {A Survey on Evaluation of Large Language Models},
year = {2024},
issue_date = {June 2024},
publisher = {Association for Computing Machinery},
address = {New York, NY, USA},
volume = {15},
number = {3},
issn = {2157-6904},
url = {https://doi.org/10.1145/3641289},
doi = {10.1145/3641289},
month = mar,
articleno = {39},
numpages = {45}
}

@inproceedings{zamfirescu2023johnny,
author = {Zamfirescu-Pereira, J.D. and Wong, Richmond Y. and Hartmann, Bjoern and Yang, Qian},
title = {Why Johnny Can’t Prompt: How Non-AI Experts Try (and Fail) to Design LLM Prompts},
year = {2023},
isbn = {9781450394215},
publisher = {Association for Computing Machinery},
address = {New York, NY, USA},
url = {https://doi.org/10.1145/3544548.3581388},
doi = {10.1145/3544548.3581388},
articleno = {437},
numpages = {21},
location = {Hamburg, Germany},
series = {CHI '23}
}

@article{barman2024beyond,
  title={Beyond transparency and explainability: on the need for adequate and contextualized user guidelines for LLM use},
  author={Barman, Kristian Gonz{\'a}lez and Wood, Nathan and Pawlowski, Pawel},
  journal={Ethics and Information Technology},
  volume={26},
  number={3},
  pages={47},
  year={2024},
  publisher={Springer}
}

@inproceedings{bansal2019beyond,
  title={Beyond accuracy: The role of mental models in human-AI team performance},
  author={Bansal, Gagan and Nushi, Besmira and Kamar, Ece and Lasecki, Walter S and Weld, Daniel S and Horvitz, Eric},
  booktitle={Proceedings of the AAAI conference on human computation and crowdsourcing},
  volume={7},
  pages={2--11},
  year={2019}
}

@inproceedings{bansal2021does,
  title={Does the whole exceed its parts? the effect of ai explanations on complementary team performance},
  author={Bansal, Gagan and Wu, Tongshuang and Zhou, Joyce and Fok, Raymond and Nushi, Besmira and Kamar, Ece and Ribeiro, Marco Tulio and Weld, Daniel},
  booktitle={Proceedings of the 2021 CHI conference on human factors in computing systems},
  pages={1--16},
  year={2021}
}

@inproceedings{redmiles2016learned,
  title={How i learned to be secure: a census-representative survey of security advice sources and behavior},
  author={Redmiles, Elissa M and Kross, Sean and Mazurek, Michelle L},
  booktitle={Proceedings of the 2016 ACM SIGSAC conference on computer and communications security},
  pages={666--677},
  year={2016}
}

@inproceedings{zhang2024s,
  title={“It's a Fair Game”, or Is It? Examining How Users Navigate Disclosure Risks and Benefits When Using LLM-Based Conversational Agents},
  author={Zhang, Zhiping and Jia, Michelle and Lee, Hao-Ping and Yao, Bingsheng and Das, Sauvik and Lerner, Ada and Wang, Dakuo and Li, Tianshi},
  booktitle={Proceedings of the CHI Conference on Human Factors in Computing Systems},
  pages={1--26},
  year={2024}
}

@inproceedings{wash2011influencing,
  title={Influencing mental models of security: a research agenda},
  author={Wash, Rick and Rader, Emilee},
  booktitle={Proceedings of the 2011 New Security Paradigms Workshop},
  pages={57--66},
  year={2011}
}

@techreport{schoeffer2024ai,
  title={AI Reliance and Decision Quality: Fundamentals, Interdependence, and the Effects of Interventions},
  author={Schoeffer, Jakob and Jakubik, Johannes and V{\"o}ssing, Michael and K{\"u}hl, Niklas and Satzger, Gerhard},
  year={2024},
  institution={Center for Open Science}
}

@article{williams2004legal,
  title={Legal English and plain language: An introduction},
  author={Williams, Christopher and others},
  journal={ESP across Cultures},
  volume={1},
  number={1},
  pages={111--124},
  year={2004}
}

@article{zeffane2011communication,
  title={Communication, commitment \& trust: Exploring the triad},
  author={Zeffane, Rachid and Tipu, Syed A and Ryan, James C},
  journal={International Journal of Business and Management},
  volume={6},
  number={6},
  pages={77--87},
  year={2011}
}

@article{flowerday-2007,
	author = {S. Flowerday and R. von Solms},
	title = {What constitutes information integrity?},
	journal = {South African Journal of Information Management},
	volume = {9},
	number = {4},
	year = {2007},
	issn = {1560-683X},	doi = {10.4102/sajim.v9i4.201},
	url = {https://sajim.co.za/index.php/sajim/article/view/201}
}

@article{blooms-taxonomy-2001,
 ISSN = {0013175X, 21623163},
 URL = {http://www.jstor.org/stable/42926529},
 author = {Jack Conklin},
 journal = {Educational Horizons},
 number = {3},
 pages = {154--159},
 publisher = {[Sage Publications, Ltd., Phi Delta Kappa International]},
 reviewed-author = {Lorin W. Anderson and David Krathwohl and Peter Airasian and Kathleen A. Cruikshank and Richard E. Mayer and Paul Pintrich and James Raths and Merlin C. Wittrock},
 urldate = {2024-09-04},
 volume = {83},
 year = {2005}
}

@inproceedings {redmiles2020webadvice,
author = {Elissa M. Redmiles and Noel Warford and Amritha Jayanti and Aravind Koneru and Sean Kross and Miraida Morales and Rock Stevens and Michelle L. Mazurek},
title = {A Comprehensive Quality Evaluation of Security and Privacy Advice on the Web},
booktitle = {29th USENIX Security Symposium (USENIX Security 20)},
year = {2020},
isbn = {978-1-939133-17-5},
pages = {89--108},
url = {https://www.usenix.org/conference/usenixsecurity20/presentation/redmiles},
publisher = {USENIX Association},
month = aug,
}

@article{robert-2017-152steps,
title	= {152 Simple Steps to Stay Safe Online: Security Advice for Non-Tech-Savvy Users},
author	= {Robert W. Reeder and Iulia Ion and Sunny Consolvo},
year	= {2017},
journal	= {IEEE Security and Privacy}
}

@article{sruti-2022-ambiguous-sec-advice,
author = {Bhagavatula, Sruti and Bauer, Lujo and Kapadia, Apu},
title = {"Adulthood is Trying Each of the Same Six Passwords That You Use for Everything": The Scarcity and Ambiguity of Security Advice on Social Media},
year = {2022},
issue_date = {November 2022},
publisher = {Association for Computing Machinery},
address = {New York, NY, USA},
volume = {6},
number = {CSCW2},
url = {https://doi.org/10.1145/3555154},
doi = {10.1145/3555154},
journal = {Proc. ACM Hum.-Comput. Interact.},
month = {11},
articleno = {264},
numpages = {27}
}

@inproceedings {hasegawa-2022-non-experts-snp-qna,
author = {Ayako A. Hasegawa and Naomi Yamashita and Tatsuya Mori and Daisuke Inoue and Mitsuaki Akiyama},
title = {Understanding {Non-Experts{\textquoteright}} Security- and {Privacy-Related} Questions on a {Q\&A} Site},
booktitle = {Eighteenth Symposium on Usable Privacy and Security (SOUPS 2022)},
year = {2022},
isbn = {978-1-939133-30-4},
address = {Boston, MA},
pages = {39--56},
url = {https://www.usenix.org/conference/soups2022/presentation/hasegawa},
publisher = {USENIX Association},
month = aug,
}

@article{pattnaik2023perspectives,
  title={Perspectives of non-expert users on cyber security and privacy: An analysis of online discussions on twitter},
  author={Pattnaik, Nandita and Li, Shujun and Nurse, Jason RC},
  journal={Computers \& Security},
  volume={125},
  pages={103008},
  year={2023},
  publisher={Elsevier}
}

@inproceedings{redmiles-2016-learned-to-be-secure,
author = {Redmiles, Elissa M. and Kross, Sean and Mazurek, Michelle L.},
title = {How I Learned to Be Secure: A Census-Representative Survey of Security Advice Sources and Behavior},
year = {2016},
isbn = {9781450341394},
publisher = {Association for Computing Machinery},
address = {New York, NY, USA},
url = {https://doi.org/10.1145/2976749.2978307},
doi = {10.1145/2976749.2978307},
booktitle = {Proceedings of the 2016 ACM SIGSAC Conference on Computer and Communications Security},
pages = {666–677},
numpages = {12},
location = {Vienna, Austria},
series = {CCS '16}
}

@inproceedings {nicholson-2018-cybersurvival,
author = {James Nicholson and Lynne Coventry and Pam Briggs},
title = {Introducing the Cybersurvival Task: Assessing and Addressing Staff Beliefs about Effective Cyber Protection},
booktitle = {Fourteenth Symposium on Usable Privacy and Security (SOUPS 2018)},
year = {2018},
isbn = {978-1-939133-10-6},
address = {Baltimore, MD},
pages = {443--457},
url = {https://www.usenix.org/conference/soups2018/presentation/nicholson},
publisher = {USENIX Association},
month = aug,
}

@article{reeder-2017-152-steps,
title	= {152 Simple Steps to Stay Safe Online: Security Advice for Non-Tech-Savvy Users},
author	= {Robert W. Reeder and Iulia Ion and Sunny Consolvo},
year	= {2017},
journal	= {IEEE Security and Privacy}
}

@INPROCEEDINGS{redmiles-2016-sources-n-selection,
  author={Redmiles, Elissa M. and Malone, Amelia R. and Mazurek, Michelle L.},
  booktitle={2016 IEEE Symposium on Security and Privacy (SP)}, 
  title={I Think They're Trying to Tell Me Something: Advice Sources and Selection for Digital Security}, 
  year={2016},
  volume={},
  number={},
  pages={272-288},
  doi={10.1109/SP.2016.24}}

@inproceedings{emilee-2012-informal-lessons,
author = {Rader, Emilee and Wash, Rick and Brooks, Brandon},
title = {Stories as Informal Lessons about Security},
year = {2012},
isbn = {9781450315326},
publisher = {Association for Computing Machinery},
address = {New York, NY, USA},
url = {https://doi.org/10.1145/2335356.2335364},
doi = {10.1145/2335356.2335364},
booktitle = {Proceedings of the Eighth Symposium on Usable Privacy and Security},
articleno = {6},
numpages = {17},
location = {Washington, D.C.},
series = {SOUPS '12}
}

@inproceedings {pfeffer-2022-replication-infomal-lessons,
author = {Katharina Pfeffer and Alexandra Mai and Edgar Weippl and Emilee Rader and Katharina Krombholz},
title = {Replication: Stories as Informal Lessons about Security},
booktitle = {Eighteenth Symposium on Usable Privacy and Security (SOUPS 2022)},
year = {2022},
isbn = {978-1-939133-30-4},
address = {Boston, MA},
pages = {1--18},
url = {https://www.usenix.org/conference/soups2022/presentation/pfeffer},
publisher = {USENIX Association},
month = aug,
}

@inproceedings {ion-2015-expert-v-non-expert,
author = {Iulia Ion and Rob Reeder and Sunny Consolvo},
title = {{{\textquotedblleft}...No} one Can Hack My {Mind{\textquotedblright}}: Comparing Expert and {Non-Expert} Security Practices},
booktitle = {Eleventh Symposium On Usable Privacy and Security (SOUPS 2015)},
year = {2015},
isbn = {978-1-931971-249},
address = {Ottawa},
pages = {327--346},
url = {https://www.usenix.org/conference/soups2015/proceedings/presentation/ion},
publisher = {USENIX Association},
month = jul,
}

@inproceedings {busse-2019-replication-expert-v-non-expert,
author = {Karoline Busse and Julia Sch{\"a}fer and Matthew Smith},
title = {Replication: No One Can Hack My Mind Revisiting a Study on Expert and {Non-Expert} Security Practices and Advice},
booktitle = {Fifteenth Symposium on Usable Privacy and Security (SOUPS 2019)},
year = {2019},
isbn = {978-1-939133-05-2},
address = {Santa Clara, CA},
pages = {117--136},
url = {https://www.usenix.org/conference/soups2019/presentation/busse},
publisher = {USENIX Association},
month = aug,
}

@article{murray-2023-costbenefitsofadvice,
author = {Murray, Hazel and Malone, David},
title = {Costs and Benefits of Authentication Advice},
year = {2023},
issue_date = {August 2023},
publisher = {Association for Computing Machinery},
address = {New York, NY, USA},
volume = {26},
number = {3},
issn = {2471-2566},
url = {https://doi.org/10.1145/3588031},
doi = {10.1145/3588031},
journal = {ACM Trans. Priv. Secur.},
month = {05},
articleno = {30},
numpages = {35}
}

@inproceedings {fulton-2019-media-effect,
author = {Kelsey R. Fulton and Rebecca Gelles and Alexandra McKay and Yasmin Abdi and Richard Roberts and Michelle L. Mazurek},
title = {The Effect of Entertainment Media on Mental Models of Computer Security},
booktitle = {Fifteenth Symposium on Usable Privacy and Security (SOUPS 2019)},
year = {2019},
isbn = {978-1-939133-05-2},
address = {Santa Clara, CA},
pages = {79--95},
url = {https://www.usenix.org/conference/soups2019/presentation/fulton},
publisher = {USENIX Association},
month = aug,
}

@article{shay-2016-password-policy,
author = {Shay, Richard and Komanduri, Saranga and Durity, Adam L. and Huh, Phillip (Seyoung) and Mazurek, Michelle L. and Segreti, Sean M. and Ur, Blase and Bauer, Lujo and Christin, Nicolas and Cranor, Lorrie Faith},
title = {Designing Password Policies for Strength and Usability},
year = {2016},
issue_date = {May 2016},
publisher = {Association for Computing Machinery},
address = {New York, NY, USA},
volume = {18},
number = {4},
issn = {1094-9224},
url = {https://doi.org/10.1145/2891411},
doi = {10.1145/2891411},
month = {05},
articleno = {13},
numpages = {34}
}

@inproceedings {habib-2018-password-expiration,
author = {Hana Habib and Pardis Emami Naeini and Summer Devlin and Maggie Oates and Chelse Swoopes and Lujo Bauer and Nicolas Christin and Lorrie Faith Cranor},
title = {User Behaviors and Attitudes Under Password Expiration Policies},
booktitle = {Fourteenth Symposium on Usable Privacy and Security (SOUPS 2018)},
year = {2018},
isbn = {978-1-939133-10-6},
address = {Baltimore, MD},
pages = {13--30},
url = {https://www.usenix.org/conference/soups2018/presentation/habib-password},
publisher = {USENIX Association},
month = aug
}

@inproceedings{zhang-2010-password-expiration,
author = {Zhang, Yinqian and Monrose, Fabian and Reiter, Michael K.},
title = {The security of modern password expiration: an algorithmic framework and empirical analysis},
year = {2010},
isbn = {9781450302456},
publisher = {Association for Computing Machinery},
address = {New York, NY, USA},
url = {https://doi.org/10.1145/1866307.1866328},
doi = {10.1145/1866307.1866328},
booktitle = {Proceedings of the 17th ACM Conference on Computer and Communications Security},
pages = {176–186},
numpages = {11},
location = {Chicago, Illinois, USA},
series = {CCS '10}
}

@article{chiasson-2015-password-expiration,
author = {Chiasson, Sonia and Oorschot, P. C.},
title = {Quantifying the security advantage of password expiration policies},
year = {2015},
issue_date = {December  2015},
publisher = {Kluwer Academic Publishers},
address = {USA},
volume = {77},
number = {2–3},
issn = {0925-1022},
url = {https://doi.org/10.1007/s10623-015-0071-9},
doi = {10.1007/s10623-015-0071-9},
journal = {Des. Codes Cryptography},
month = {12},
pages = {401–408},
numpages = {8}
}

@misc{grassi-2017-nist-password-guidelines,
  author = {Paul Grassi and Ray Perlner and Elaine Newton and Andrew Regenscheid and William Burr and Justin Richer and Naomi Lefkovitz and Jamie Danker and Mary Theofanos},
  title = {Digital Identity Guidelines: Authentication and Lifecycle Management [including updates as of 12- 01-2017]},
  year = {2017},
  month = {12},
  publisher = {Special Publication (NIST SP), National Institute of Standards and Technology, Gaithersburg, MD},
  doi = {https://doi.org/10.6028/NIST.SP.800-63b},
  language = {en},
}

@inproceedings{dekoven-security-practices-2019,
author = {DeKoven, Louis F. and Randall, Audrey and Mirian, Ariana and Akiwate, Gautam and Blume, Ansel and Saul, Lawrence K. and Schulman, Aaron and Voelker, Geoffrey M. and Savage, Stefan},
title = {Measuring Security Practices and How They Impact Security},
year = {2019},
isbn = {9781450369480},
publisher = {Association for Computing Machinery},
address = {New York, NY, USA},
url = {https://doi.org/10.1145/3355369.3355571},
doi = {10.1145/3355369.3355571},
booktitle = {Proceedings of the Internet Measurement Conference},
pages = {36–49},
numpages = {14},
location = {Amsterdam, Netherlands},
series = {IMC '19}
}

@inproceedings{girish-local-iot-2023,
author = {Girish, Aniketh and Hu, Tianrui and Prakash, Vijay and Dubois, Daniel J. and Matic, Srdjan and Huang, Danny Yuxing and Egelman, Serge and Reardon, Joel and Tapiador, Juan and Choffnes, David and Vallina-Rodriguez, Narseo},
title = {In the Room Where It Happens: Characterizing Local Communication and Threats in Smart Homes},
year = {2023},
isbn = {9798400703829},
publisher = {Association for Computing Machinery},
address = {New York, NY, USA},
url = {https://doi.org/10.1145/3618257.3624830},
doi = {10.1145/3618257.3624830},
booktitle = {Proceedings of the 2023 ACM on Internet Measurement Conference},
pages = {437–456},
numpages = {20},
location = {Montreal QC, Canada},
series = {IMC '23}
}

@misc{cert_vu240311,
  author       = {{CERT Coordination Center}},
  title        = {Multiple Bluetooth implementation vulnerabilities affect many devices - Vulnerability Note VU\#240311},
  howpublished = {\url{https://www.kb.cert.org/vuls/id/240311}},
  year         = {2017},
  note         = {Original Release Date: 2017-09-12, Last Revised: 2017-11-08},
  url          = {https://www.kb.cert.org/vuls/id/240311}
}

@book{zinsser2001writing,
  title={On writing well: The classic guide to writing nonfiction},
  author={Zinsser, William Knowlton},
  year={2001},
  publisher={Quill/A Harper Collins Books}
}

@inproceedings{chen2023can,
  title={Can large language models provide security \& privacy advice? measuring the ability of llms to refute misconceptions},
  author={Chen, Yufan and Arunasalam, Arjun and Celik, Z Berkay},
  booktitle={Proceedings of the 39th Annual Computer Security Applications Conference},
  pages={366--378},
  year={2023}
}

@book{solnit2014men,
  title={Men explain things to me},
  author={Solnit, Rebecca},
  year={2014},
  publisher={Haymarket books}
}

@article{next_web,
author={Ahlgren, Linnea},
title={Meet Finn — bunq’s new GenAI chatbot},
journal={The Next Web, December 19},
url={https://thenextweb.com/news/bunq-new-generative-ai-chatbot-finn},
year={2023}
}

@article{zdnet_norton,
title={How to use Norton's free ai-powered scam detector},
author={Whitney, Lance},
journal={ZDNet, September 14},
url={https://www.zdnet.com/article/how-to-use-nortons-free-ai-powered-scam-detector/},
year={2023}
}

@article{biocatch,
author={Ruden, Seth},
title={How to Level Up Your Fraud Defense with ChatGPT's Own AI Suggestions},
journal={BioCatch Blog Channel},
url={https://www.biocatch.com/blog/chatgpt-improve-fraud-detection},
year={2023},
month={01},
day={27}
}

@article{dataleon,
author={Sarah},
title={Using ChatGPT for real-time fraud detection in finance},
journal={Dataleon. Ai Blog, December 11},
year={2023},
url={https://www.dataleon.ai/en/blog/using-chatgpt-for-real-time-fraud-detection-in-finance},
}

@inproceedings{kim-https-phishing-CA-2021,
author = {Kim, Doowon and Cho, Haehyun and Kwon, Yonghwi and Doup\'{e}, Adam and Son, Sooel and Ahn, Gail-Joon and Dumitras, Tudor},
title = {Security Analysis on Practices of Certificate Authorities in the HTTPS Phishing Ecosystem},
year = {2021},
isbn = {9781450382878},
publisher = {Association for Computing Machinery},
address = {New York, NY, USA},
url = {https://doi.org/10.1145/3433210.3453100},
doi = {10.1145/3433210.3453100},
booktitle = {Proceedings of the 2021 ACM Asia Conference on Computer and Communications Security},
pages = {407–420},
numpages = {14},
location = {Virtual Event, Hong Kong},
series = {ASIA CCS '21}
}

@inproceedings{felt-permissions-2011,
author = {Felt, Adrienne Porter and Chin, Erika and Hanna, Steve and Song, Dawn and Wagner, David},
title = {Android permissions demystified},
year = {2011},
isbn = {9781450309486},
publisher = {Association for Computing Machinery},
address = {New York, NY, USA},
url = {https://doi.org/10.1145/2046707.2046779},
doi = {10.1145/2046707.2046779},
booktitle = {Proceedings of the 18th ACM Conference on Computer and Communications Security},
pages = {627–638},
numpages = {12},
location = {Chicago, Illinois, USA},
series = {CCS '11}
}

@inproceedings{chia-permission-risks-2012,
author = {Chia, Pern Hui and Yamamoto, Yusuke and Asokan, N.},
title = {Is this app safe? a large scale study on application permissions and risk signals},
year = {2012},
isbn = {9781450312295},
publisher = {Association for Computing Machinery},
address = {New York, NY, USA},
url = {https://doi.org/10.1145/2187836.2187879},
doi = {10.1145/2187836.2187879},
pages = {311–320},
numpages = {10},
location = {Lyon, France},
series = {WWW '12}
}

@inproceedings{calciati-auto-permissions-2020,
author = {Calciati, Paolo and Kuznetsov, Konstantin and Gorla, Alessandra and Zeller, Andreas},
title = {Automatically Granted Permissions in Android apps: An Empirical Study on their Prevalence and on the Potential Threats for Privacy},
year = {2020},
isbn = {9781450375177},
publisher = {Association for Computing Machinery},
address = {New York, NY, USA},
url = {https://doi.org/10.1145/3379597.3387469},
doi = {10.1145/3379597.3387469},
booktitle = {Proceedings of the 17th International Conference on Mining Software Repositories},
pages = {114–124},
numpages = {11},
location = {Seoul, Republic of Korea},
series = {MSR '20}
}

@article{farquhar2024detecting,
  title={Detecting hallucinations in large language models using semantic entropy},
  author={Farquhar, Sebastian and Kossen, Jannik and Kuhn, Lorenz and Gal, Yarin},
  journal={Nature},
  volume={630},
  number={8017},
  pages={625--630},
  year={2024},
  publisher={Nature Publishing Group UK London}
}

@inproceedings{binkhorst2022security,
  title={Security at the End of the Tunnel: The Anatomy of $\{$VPN$\}$ Mental Models Among Experts and $\{$Non-Experts$\}$ in a Corporate Context},
  author={Binkhorst, Veroniek and Fiebig, Tobias and Krombholz, Katharina and Pieters, Wolter and Labunets, Katsiaryna},
  booktitle={31st USENIX Security Symposium (USENIX Security 22)},
  pages={3433--3450},
  year={2022}
}

@inproceedings{dutkowska2022and,
  title={How and why people use virtual private networks},
  author={Dutkowska-Zuk, Agnieszka and Hounsel, Austin and Morrill, Amy and Xiong, Andre and Chetty, Marshini and Feamster, Nick},
  booktitle={31st USENIX Security Symposium (USENIX Security 22)},
  pages={3451--3465},
  year={2022}
}

@inproceedings{herbert2023world,
  title={A world full of privacy and security (mis) conceptions? Findings of a representative survey in 12 countries},
  author={Herbert, Franziska and Becker, Steffen and Schaewitz, Leonie and Hielscher, Jonas and Kowalewski, Marvin and Sasse, Angela and Acar, Yasemin and D{\"u}rmuth, Markus},
  booktitle={Proceedings of the 2023 CHI Conference on Human Factors in Computing Systems},
  pages={1--23},
  year={2023}
}

@inproceedings{wash2010folk,
  title={Folk models of home computer security},
  author={Wash, Rick},
  booktitle={Proceedings of the Sixth Symposium on Usable Privacy and Security},
  pages={1--16},
  year={2010}
}
\appendix

\if \conference1
    \section{Validation of LLM Response Reliability} \label{appendix-repro-validation}
\fi

\if \conference2
    \subsection{Validation of LLM Response Reliability} \label{appendix-repro-validation}
\fi

\begin{table}[!h]

\if \conference1
\relsize{-1}
\fi

    \centering
    \setlength{\abovecaptionskip}{0pt}
    \caption{Evaluation results of paraphrasing experiment where 10 questions were paraphrased 42 times.}
    \label{tab:prompt-sensitivity-eval}
    \begin{tabular}{p{0.11\linewidth} p{0.16\linewidth} p{0.12\linewidth} p{0.18\linewidth} p{0.15\linewidth}}
        \toprule
        Model & Completely similar & Mostly similar & Insufficiently similar & Consistent? \\
        \midrule
        GPT & 2 & 37 & 3 & Yes\\
        Llama & 0 & 38 & 4  & Yes\\
        Gemini & 0  & 35  & 7  & Yes\\
        \bottomrule
    \end{tabular}
\end{table}

\paragraph{Are LLM responses consistent across paraphrased questions?} Our experiment comprised 10 questions paraphrased to generate 42 additional questions. GPT, Llama, and Gemini generated 52 responses each for these questions. The breakdown of our assessment is in the table \ref{tab:prompt-sensitivity-eval}. Based on our evaluation, we concluded all the models generate mostly similar responses to semantically similar prompts, so LLM responses are consistent for our purposes, even when a few responses for all models were evaluated as insufficiently similar, all the researchers as a group concluded that they were not different enough to change our conclusion. 

\paragraph{Takeaway.} LLM responses are mostly insensitive to paraphrasing of open-ended user security questions. As such, for each of our security questions, we only asked an LLM once without rewording.

\begin{table}[!h]

\if \conference1
\relsize{-1}
\fi
    \centering
    \setlength{\abovecaptionskip}{0pt}
    \caption{Evaluation results of repetitive consistency experiment where 7 questions were asked repeatedly 5 times.}
    \label{tab:repetitive-consistency-eval}
    \begin{tabular}{p{0.11\linewidth} p{0.16\linewidth} p{0.12\linewidth} p{0.18\linewidth} p{0.18\linewidth}}
        \toprule
        Model & Completely similar & Mostly similar & Insufficiently similar & Consistent? \\
        \midrule
        GPT & 0 & 6 & 1 & Yes\\
        Llama & 0 & 7 & 0 & Yes\\
        Gemini & 0 & 7 & 0  & Yes\\
        \bottomrule
    \end{tabular}
\end{table}

\paragraph{Are LLM responses consistent across repeated questions?} Our experiment comprised 7 questions, where each question was repeatedly used to generate 5 responses for all three LLMs, 35 responses in total from each model. The breakdown of our assessment is in the table \ref{tab:repetitive-consistency-eval}. From this experiment, we concluded that all three LLMs in our evaluation generate mostly similar responses for a question asked repeatedly, thus their responses have consistency.

\paragraph{Takeaway.} LLM responses are mostly consistent given the same repeated questions. As such, for each of our security questions, we only asked an LLM once without repetition.

\if \conference1
    \section{Heuristics for Users} \label{gpt-eval-results}
\fi

\if \conference2
    \subsection{Heuristics for Users} \label{gpt-eval-results}
\fi

To give users simple heuristics to detect errors in LLM responses, we calculated the conditional probability of error in other evaluation criteria given that the user sees an error in one evaluation criterion (e.g., Directness), and present it in \cref{appendix-tab:cond-probabilities}.

\begin{table}[!h]
    \setlength{\abovecaptionskip}{0pt}
\caption{Given error in one criterion, conditional probabilities of error rates in other evaluation criteria in GPT responses.}
\label{appendix-tab:cond-probabilities}
\if \conference1
\relsize{-1}
\fi
\resizebox{\linewidth}{!}{%
    \begin{tabular}{p{0.2\linewidth}p{0.20\linewidth}p{0.22\linewidth}p{0.225\linewidth}}
    \toprule
    Given error & \# of responses & \# of err in others & $P($error$)$ in others \\
    \midrule
    Not Correct & 246 & 139 & 0.565 \\
    Not Thorough & 287 & 154 & 0.537 \\
    Not Relevant & 15 & 11 & 0.733 \\
    Not Direct & 153 & 90 & 0.588 \\
    \bottomrule
    \end{tabular}
}
\end{table}

\begin{table*}[!hb]
\if \conference1
\relsize{-1}
\fi
    \setlength{\abovecaptionskip}{0pt}
\caption{Detailed performance of GPT in terms of \% of responses where it fully achieved the indicated attribute.}
\label{tab:appendix-gpt-evaluation-dist-big-with-directness}

\resizebox{\textwidth}{!}{%
    {\renewcommand{\arraystretch}{1.2}%
    \begin{tabular}{p{0.07\linewidth}p{0.2\linewidth}p{0.04\linewidth}p{0.06\linewidth}p{0.05\linewidth}p{0.05\linewidth}p{0.06\linewidth}p{0.06\linewidth}p{0.04\linewidth}p{0.06\linewidth}p{0.05\linewidth}p{0.04\linewidth}p{0.04\linewidth}}
    \toprule
     &  & \multicolumn{3}{l}{Accuracy} & \multicolumn{3}{l}{Thoroughness} & \multicolumn{3}{l}{Relevancy} & \multicolumn{2}{l}{Directness}\\
     \cmidrule(lr){3-5} \cmidrule(lr){6-8} \cmidrule(lr){9-11} \cmidrule(lr){12-13}
    {Categorization} &\ Categories & Correct & Somewhat correct & Not correct & Thorough & Somewhat thorough & Not thorough & Relevant & Somewhat relevant & Not relevant & Direct & Not direct \\
    \midrule
    \textbf{} & \textbf{\# of all questions (900)} & 0.73 & 0.26 & 0.01 & 0.68 & 0.30 & 0.02 & 0.98 & 0.01 & 0.01 & 0.83 & 0.17 \\
    \cline{1-13}
    \multirow[t]{7}{\linewidth}{\textbf{Security Category}} & \textbf{Authentication (449)} & 0.69 & 0.30 & 0.01 & 0.69 & 0.29 & 0.02 & 0.98 & 0.01 & 0.01 & 0.83 & 0.17 \\
    \textbf{} & \textbf{Scams (83)} & 0.82 & 0.18 & 0.00 & 0.70 & 0.30 & 0.00 & 1.00 & 0.00 & 0.00 & 0.83 & 0.17 \\
    \textbf{} & \textbf{Safe browsing (50)} & 0.80 & 0.20 & 0.00 & 0.68 & 0.28 & 0.04 & 0.96 & 0.04 & 0.00 & 0.88 & 0.12 \\
    \textbf{} & \textbf{Av (163)} & 0.61 & 0.37 & 0.02 & 0.59 & 0.39 & 0.02 & 0.99 & 0.00 & 0.01 & 0.86 & 0.14 \\
    \textbf{} & \textbf{Secure network/wifi (46)} & 0.87 & 0.13 & 0.00 & 0.67 & 0.33 & 0.00 & 0.96 & 0.04 & 0.00 & 0.70 & 0.30 \\
    \textbf{} & \textbf{Smart devs (19)} & 0.89 & 0.11 & 0.00 & 0.63 & 0.26 & 0.11 & 0.95 & 0.00 & 0.05 & 0.79 & 0.21 \\
    \textbf{} & \textbf{Updates (90)} & 0.88 & 0.12 & 0.00 & 0.81 & 0.18 & 0.01 & 0.99 & 0.01 & 0.00 & 0.84 & 0.16 \\
    \cline{1-13}
    \multirow[t]{5}{\linewidth}{\textbf{Themes of questions}} & \textbf{Glossary and Facts (355)} & 0.69 & 0.30 & 0.01 & 0.75 & 0.23 & 0.02 & 0.98 & 0.02 & 0.01 & 0.85 & 0.15 \\
    \textbf{} & \textbf{Guidance and Best Practices (399)} & 0.72 & 0.27 & 0.01 & 0.61 & 0.37 & 0.02 & 0.99 & 0.01 & 0.01 & 0.81 & 0.19 \\
    \textbf{} & \textbf{Trends (17)} & 0.47 & 0.53 & 0.00 & 0.82 & 0.18 & 0.00 & 1.00 & 0.00 & 0.00 & 0.76 & 0.24 \\
    \textbf{} & \textbf{Risk Assessment (100)} & 0.85 & 0.13 & 0.02 & 0.68 & 0.29 & 0.03 & 0.97 & 0.02 & 0.01 & 0.88 & 0.12 \\
    \textbf{} & \textbf{Attacks (29)} & 1.00 & 0.00 & 0.00 & 0.83 & 0.17 & 0.00 & 1.00 & 0.00 & 0.00 & 0.72 & 0.28 \\
    \cline{1-13}
    \multirow[t]{3}{\linewidth}{\textbf{Knowledge}} & \textbf{Factual (437)} & 0.76 & 0.23 & 0.01 & 0.72 & 0.25 & 0.03 & 0.98 & 0.01 & 0.01 & 0.84 & 0.16 \\
    \textbf{} & \textbf{Conceptual (164)} & 0.72 & 0.28 & 0.00 & 0.77 & 0.23 & 0.01 & 0.98 & 0.02 & 0.00 & 0.89 & 0.11 \\
    \textbf{} & \textbf{Procedural (299)} & 0.69 & 0.30 & 0.01 & 0.57 & 0.41 & 0.02 & 0.99 & 0.01 & 0.00 & 0.78 & 0.22 \\
    \bottomrule
    \end{tabular}
    }
    }

\end{table*}

\if \conference1
    \section{More Error Themes and Patterns in Models} \label{appendix-error-patterns}
\fi

\if \conference2
    \subsection{More Error Themes and Patterns in Models} \label{appendix-error-patterns}
\fi

\if \conference1
    \subsection{\Permissions} 
\fi

\if \conference2
    \subsubsection{\Permissions} 
\fi
    All the LLMs in our evaluation omitted and listed incorrect mobile permissions used by different security products, like AV, password manager, and MFA applications, and sometimes failed to explain why permission was used on mobiles or computers. They often mentioned AVs needed camera, microphone, phone, and SMS permissions, but these are not listed in their official documentation. When explaining why permissions like \textit{location} and \textit{nativeMessaging} were used by MFA mobile applications and browser extensions, their responses failed to mention the reason is access control and to support biometrics.

\if \conference1
    \subsection{\Scams}
\fi

\if \conference2
    \subsubsection{\Scams}
\fi
    All the LLMs were not completely thorough in explaining scams in terms of what exactly a scam is about, who the target audience is and who reports more, their recent trends and the consequences. When explaining cryptocurrency scams, they didn't mention these scams ask anyone to invest money and do not just target people with cryptocurrencies. GPT failed to explain the AV software scam. When asked about the target audience and report, they conflated targeting and reporting and didn't mention that the elderly population (65+) gets targeted more and young adults (18 - 65) report more scams \cite{ftc_scams_all_ages_2022}. Llama and GPT didn't mention that the underage (\textless18) population also experiences and reports scams \cite{ftc_consumer_sentinal_scam_by_age_2023}. No models mentioned that scams could lead to computational resource misuse, not just financial loss.

\if \conference1
    \subsection{\MakingUpFactsAV}
\fi

\if \conference2
    \subsubsection{\MakingUpFactsAV}
\fi
    We found that LLMs frequently generated fabricated information when discussing the capabilities of antivirus (AV) software. For example, they incorrectly claimed that AVs are available for IoT devices, NAS systems, and gaming consoles. GPT invented the term “file emulation” to explain AV functionality, while both GPT and Llama made unverifiable claims, such as AVs being inherently “lightweight,” “efficient,” and not affecting battery life—claims contradicted by user reports on forums like Reddit.

    Llama also fabricated specific statistics about battery consumption by various AVs and made misleading comparisons to activities like browsing Facebook or playing Candy Crush. Across our dataset, LLMs exaggerated AV features not listed on official websites, such as call blockers and anti-theft tools. Regarding ransomware protection, LLMs consistently claimed AVs detect suspicious activity in real time and can decrypt encrypted files—claims inconsistent with AVs' actual permission-based block lists and lack of file recovery capabilities. Notably, Gemini mischaracterized AVG’s Hack Alerts as a scam or phishing attempt.

\if \conference1
    \subsection{\TemporalDependency}
\fi

\if \conference2
    \subsubsection{\TemporalDependency}
\fi
    All the models, but more specifically GPT, showed temporal dependency unawareness, where they would present solutions that can only be taken retroactively. They all suggested backing up data after malware infection, and taking action on the lost phone used for 2FA. GPT suggested revoking security keys after they are lost, which won't be possible without recovering the account; using Okta TOTP in case the mobile is inaccessible; downloading or updating AVs after disconnecting the device from the internet in case of malware infection.

\if \conference1
    \subsection{\IncorrectURLsEmails}
\fi

\if \conference2
    \subsubsection{\IncorrectURLsEmails}
\fi
     All the models confidently linked URLs and email addresses in their responses. URLs were either inaccessible or landed on random websites, and emails were not active email addresses or non-existent. We saw this behavior 11 times in GPT across all 900 questions, and 7 and 5 times in Llama and Gemini, respectively, in our limited 304-response evaluation. 
\if \conference1
    \subsection{\ResponsesGearedTowardsDevelopers.}
\fi

\if \conference2
    \subsubsection{\ResponsesGearedTowardsDevelopers.}
\fi
    Often answering basic security questions, the LLMs we studied explained them with too much technical content that is unlikely to be interpretable by end users. For the question “How do antivirus work?", below is an excerpt from Gemini's response:

    \begin{quote}
    \uline{Heuristic Analysis}: This method looks for suspicious patterns in code that may indicate malicious behavior, even if the program hasn't been seen before. It's like looking for signs of a virus based on how it acts, not just what it looks like.
    \end{quote}
    \begin{quote}
    \uline{Machine Learning}: More advanced antivirus programs use AI to analyze large amounts of data and learn to identify new malware threats. They can detect anomalies and unusual behavior, even if they haven't encountered that specific threat before.
    \end{quote}

\end{document}